# Nano-gap electrode dielectrophoresis for tether-free trapping and interferometric-scattering detection of single 20 nm particles

*Jacco Ton‡, Théo Travers‡, Jamal Soltani, Daniel Wijnperle, Dmytro Shavlovskyi, Michel Orrit and Sergii Pud*

KEYWORDS: iSCAT, dielectrophoresis, nano-gap, DEP trapping, nanoparticles, label-free detection

## Abstract

Accurate detection and characterization of nanoparticles within confined spaces is crucial for applications ranging from nanofluidics to biotechnology. We present a novel approach that combines interferometric scattering (iSCAT) detection with trapping by dielectrophoresis (DEP) to achieve label-free detection of nanoparticles that are trapped and/or actuated between nano-gap electrodes. DEP utilizes the interaction between the induced dipole of the particle and the applied electric field to create a trapping potential. We demonstrate our method by trapping and label-free detection of down to 20 nm polystyrene nanoparticles. Additionally, we demonstrate that the signal-to-noise ratio of our detection can be boosted up to 20-fold by periodic actuation of the nanoparticle in the trap. This is done by a digital lock-in detection scheme on the modulated scattering signal. Our method holds promise for various applications, including assembly of nanoparticles, single-particle property analysis, and nanofluidic devices.

## Introduction

The advent of interferometric scattering (iSCAT) microscopy 20 years ago[1–3] is expanding the method toolbox for the study of nano-scale objects, be it for their detection, tracking, imaging, or for the quantification of their physical properties[4]. The unique strength of scattering detection, such as iSCAT but also dark-field detection, when compared to fluorescence, is that there is no blinking and bleaching of the scattering signal. This advantage has, for example, opened up the possibility to probe protein dynamics over more than 24 h with a plasmonic ruler[5]. On top of the endless photon budget the number of photons per second from scattering is also much larger than in fluorescence. Besides several other biological applications for iSCAT[6,7] the higher intensity has made it possible to perform tracking of a gold particle (20 nm diameter) on the cell membrane at down to microsecond time resolution with wide-field iSCAT microscopy[8]. Furthermore, due to its universal contrast mechanism iSCAT has a much broader applicability such as probing the restructuring of the electric double layer (EDL) around a small particle[9] or ion dynamics in a battery material[10]. In the past decade the sensitivity of iSCAT has reached single-protein sensitivity[11,12] and enabled protein mass quantification (mass photometry)[13]. Mass photometry has been commercialized[14] and can reach a mass resolution down to 19 kDa[13]. The beauty of mass photometry lies in its simplicity: The initially freely diffusing protein is detected at the moment of its unspecific adsorption to a glass coverslip. Still, to be able to identify such small analytes, integration times are needed of up to several hundreds of milliseconds.

To bypass the constraints of slow time resolution, posed by fluorescence detection and mass photometry for protein detection, methods that use a large signal enhancement through an optical resonance have been developed. They utilize either plasmonic near-field-enhancement[15,16], whispering-gallery-mode resonators[17] or photothermal detection[18,19]. These methods can all be used for fast detection and analysis of some additional analyte properties but only for a short time per analyte. A freely diffusing nano-scale analyte will traverse the near-field detection volume in less than[15] one microsecond which makes it impossible to probe slow dynamics of interest. A new method uses a water-filled Fabry-Pérot cavity[20] which potentially increases the observation time but makes quantifying the signal more difficult. The above problems of fast detection methods can be solved by trapping the single analytes in the detection volume.

Several single-particle trapping approaches have been developed that don't require the analyte to be coupled to a chemical tether or to undergo adsorption to a surface. The first type utilizes the passive trapping potential of an electrostatic trap[21,22] or uses geometric confinement in a nanocavity[23]. For protein-sized analytes these methods increase the observation times up to seconds. A fundamentally different approach is that of the Anti-Brownian electrokinetic (ABEL) trap[24] which is an active feedback-controlled approach that allows for arbitrarily long trapping times of molecule-scale analytes. The label-free implementation of the ABEL trap is the iSABEL trap[25] which uses iSCAT signals as feedback information. Other examples are an electro-osmotic single-particle recycling scheme[26] and a trap that uses negative dielectrophoresis to manipulate a scattering particle in a chamber between two narrow nanochannels[27]. The above trapping methods can all be implemented with single-molecule optical detection through fluorescence. However, in these traps, iSCAT detection of analytes that scatter less than a 20 nm spherical gold particle has not been demonstrated yet. There is also a detection method[28] that provides relatively long observation times for small proteins without trapping them, and detects them through iSCAT on a camera. However, this method requires a very challenging fabrication of nano-channels in glass with features down to 15 nm.

The methods that combine long-term trapping with iSCAT for optical detection[21,25–27] require a minimum scattering cross section around 0.5 nm$^2$, depending on the experimental conditions such as the laser wavelength. However, when using these trapping approaches, the position of the particle in the detection volume during iSCAT signal integration is controlled much less compared to mass photometry where the analyte is adsorbed to the coverslip and the coverslip-water interface reflection acts as a fixed and very stable reference signal. In any iSCAT experiment, the z-position of the scattering particle influences the detected intensity ($I_{det}$) through a superposition of the reference light field ($E_{ref}$) with the scattered field ($E_{scat}$) from the object in the detection volume[29,30] according to:

$$I_{det} \propto |E_{ref} + E_{scat}|^2 = |E_{ref}|^2 + |E_{scat}|^2 + 2|E_{ref}||E_{scat}|\cos(\varphi). \quad (1)$$

Especially it is important to notice that $I_{det}$ depends on the phase difference $\varphi = \varphi_{ref} - \varphi_{scat}$ between the interfering fields $E_{ref}$ and $E_{scat}$. In a standard iSCAT experiment[2] this phase term strongly depends on the difference in z-position between the scatterer and the source of the reference field (commonly the glass-water interface) and to a lesser extent on the *x* and *y* positions in the confocal detection volume. A good example of how signal fluctuations degrade trapping stability is the current iSABEL geometry[25] which only controls the x and y position very tightly while the particle can diffuse in the z-direction over a range[25] of 700 nm. When the particle diffuses in the z-direction of the trap volume, in some z-positions the signal of a small particle may be too low to properly detect the particle's position in the trap and it may therefore escape.

Near-field optical trapping[31] that combines trapping by a plasmonically enhanced optical field[32] with high-speed plasmon-enhanced detection is sensitive enough to trap and detect unlabeled proteins. However, this method suffers from ambiguity in the interpretation of the optical signal, caused by the sensitivity of the near-field detection to the exact position of the analyte in the plasmonic trapping structure. These recent developments reflect the intricacies of extremely sensitive label-free methodologies[29,33], which must provide biochemical insights about the analytes and at the same time remain accessible to a wide scientific community.

To alleviate some of the above-mentioned challenges we propose to combine confocal iSCAT detection with nano-gap dielectrophoresis (nano-DEP) as trapping method for single particles. Dielectrophoresis[34,35] is an effect that occurs when an analyte, in a solution, interacts with a spatially inhomogeneous alternating-current (AC) electric field. A dielectric particle becomes polarized through field-induced polarization, whose interaction with the field creates a trapping potential that is largest at the location with the highest field. To create this high field at a specific location in our nano-DEP trap setup (See Figure 1A as overview) the electrodes are shaped as tips. Therefore, the field is concentrated at the apex of each electrode and particles will be trapped at these positions. The use of DEP to control the position of the analyte in the detection volume bears a range of advantages: (1) it is tether-free; (2) it provides better control of the z-position of the analyte compared to the other trapping methods; (3) Compared to active feedback methods like iSABEL[25], DEP doesn't rely on detection to trap a particle; (4) DEP[36] is much more localized than electrophoresis[37], allowing for more precise spatial control of the trapping field. Therefore, it allows for the trapping of single entities with a localization precision[38] comparable to that of plasmonic trapping[32]; (5) Finally, we foresee that nano-DEP trapping can confer additional specificity to a single-particle iSCAT measurement, if the experiment will allow for optical-encoding of the information from the DEP forces. This information can be recovered through actuation as done by Ma et. Al[37], but replacing tethers by DEP interactions.

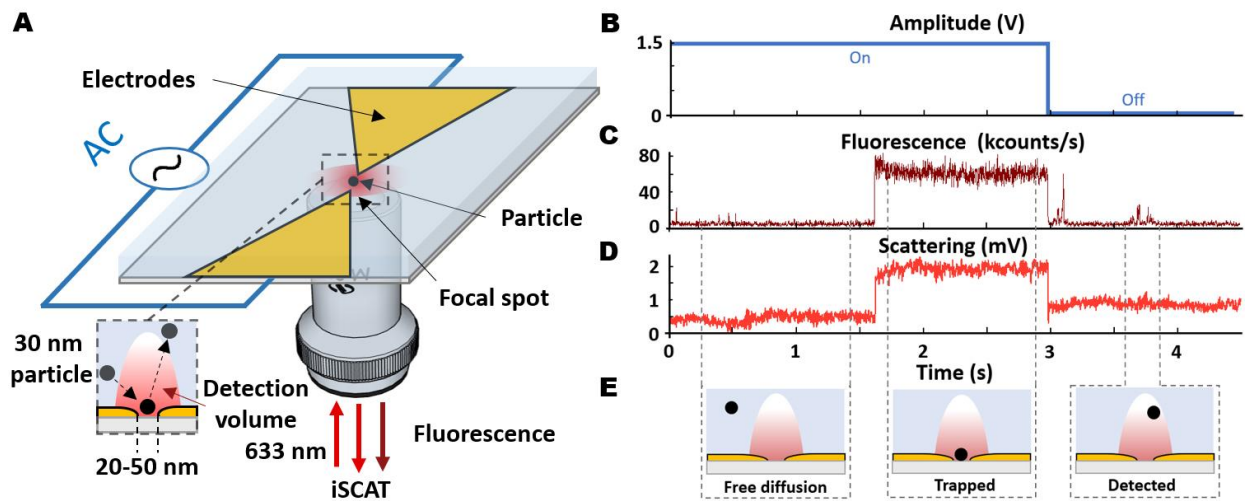

Figure 1. (A) Schematic layout of the nano-DEP and optical detection setup, with a fluorescent 30 nm particle immersed in a fluid. The fluid is rendered by a transparent blue shade above the coverslip and the particles by black dots. The red glow is the diffraction-limited focal spot of the illumination (wavelength = 633 nm). The detected wavelength in iSCAT is also 633 nm and the fluorescence is red-shifted in color. The gap between the gold electrodes (yellow) is smaller (20 to 50 nm) than the diffraction-limited spot. (B) A time trace featuring the amplitude of the AC-voltage with a frequency of 300 kHz applied over the gap between the electrodes. This AC-voltage was started earlier in time but was turned off after a certain time of having a particle trapped. (C) The corresponding fluorescence time trace before, during and after a nanoparticle is trapped in the nanogap. This fluorescence signal is relatively bright which makes us assume that this was a particle of around 30 nm from a batch of particles with nominal average diameter 20 nm. (D) Scattering time trace- of the same nanoparticle as in (C) recorded simultaneously. Notice the signal at the end of the trapping event going slightly below the baseline intensity. (E) A schematic

explanation of the particle positions when it is not detected (left), detected in both the fluorescent and scattering channel (center) and only in the fluorescence channel (right).

DEP is well established as a tool for on-chip manipulation[39,40], analysis and sorting[41–44] of microparticle ensembles. However, it has been much less widely applied to low numbers of analytes, as proposed only in a handful of publications[38,45–47]. We attribute this to the difficulty of fabrication of the rather sophisticated nanostructure designs[38,46,47] required to spatially localize a high field. Our design choice for two bow-tie-shaped tips of the electrodes (see figure 1A) was inspired by work on plasmonic trapping[48,49] and by earlier DEP implementations for the detection of surface-enhanced Raman scattering[45]. In our experiments we could reversibly trap polystyrene nanoparticles down to 20 nm diameter with DEP nanoelectrodes while applying a range of electric field parameters and detect the particles using both iSCAT and fluorescence. We have developed a feedback system based on the detection of the particle's arrival by detecting its fluorescence signal and used it to make real-time adjustments to the AC-voltage applied to our nanoelectrodes, which can be used to release the trapped particles on-demand (See figure 1B-D). Finally, we report on harmonic actuation of the trapped particle by means of periodically modulating the trapping potential. By using the modulation frequency as a reference frequency in a lock-in detection scheme we can enhance the signal-to-noise ratio of the scattering readout. We thus provide a new tool to improve the observation times of iSCAT and to acquire additional information about the analytes from the optical signal

# Results

Confining an analyte by DEP to prolong the observation time to seconds would require the trapping potential to be in the order of[27] $10\ k_B T$, where $k_B$ is the Boltzmann constant and $T$ temperature. For dielectrophoresis the trapping potential can be analytically calculated in the dipole approximation and in the limit of a linear response to the field. This approximation yields a time-averaged dielectrophoretic potential energy $\langle U_{DEP} \rangle$ in the trap for a rigid spherical particle of volume $V_p$ ($V_p = 4/3\ \pi r^3$)[34,50]:

$$\langle U_{DEP} \rangle = -\frac{3}{4} \varepsilon_0 \varepsilon_m V_p \frac{\varepsilon_p - \varepsilon_m}{\varepsilon_p + 2\varepsilon_m} |E_{AC}|^2, \qquad (2)$$

with the time-averaged force as $\langle F_{DEP} \rangle = -\nabla \langle U_{DEP} \rangle$. $\varepsilon_0$ is the absolute permittivity of vacuum and $\varepsilon_m$ ($\varepsilon_p$) the relative permittivity of the medium (particle). To prevent confusion it is important to note that here we for practical reasons have chosen to use $E_{AC}$ as the amplitude of the electric field applied over the gap between the electrodes and not the root-mean-squared of the electric field, $E_{rms}$[27,51]. In the general case of dielectrophoresis in aqueous solutions the permittivities of the particle and the medium should be regarded as complex numbers to accommodate for the losses in the respective materials. The complex permittivity $\tilde{\varepsilon}_p$ ($\tilde{\varepsilon}_m$) of the particle (medium) are $\tilde{\varepsilon}_{p,m} = \varepsilon_0 \varepsilon_{p,m} - i(\sigma_{p,m}/\omega)$, with ($\sigma_{p,m}$) – the conductivities of the two materials. This complex polarizability, $\tilde{\alpha}$, is frequency depended and is defined for a spherical particle in an AC-field, as:

$$\tilde{\alpha} = 3\varepsilon_0 \varepsilon_m V_p \left( \frac{\tilde{\varepsilon}_p - \tilde{\varepsilon}_m}{\tilde{\varepsilon}_p + 2\tilde{\varepsilon}_m} \right). \qquad (3)$$

Quantifying the energy of the interaction with the field in this case requires using the real part of the polarizability, $Re(\tilde{\alpha})$. The complex permittivity of the polystyrene particle in practice depends mainly on the surface charge and on the permittivity and conductivity of the electric double layer (EDL) that is formed on the surface of the particle. Beyond polystyrene, the trapping potential of a more complicated analyte such as a protein can be influenced if the particle possesses a static or dynamic distribution of intrinsic charges leading to a permanent dipole moment[52] or to additional surface charge effects[53]. Even if the assumptions that underpin these equations are not completely met in our nano-DEP experiment, these equations are still instructive as they show that the trapping potential will depend on particle size, on the frequency of the AC field, and on the field amplitude[54]. Due to the rapid decrease in trapping potential with size for sub-100 nm nanoparticles, maintaining a trapping potential (See equation 2) that can confine such a Brownian particle for seconds[55] requires fields as high as $10^6$-$10^7$ V/m. However, achieving such fields using conventional microscale geometries[43,56] is impossible without applying voltages that are larger than 10 V, which can lead to adverse effects, like dielectric breakdown, electrolysis, heat release, reshaping or degradation of particles, or denaturation of proteins[35]. To prevent this, we designed a nanoelectrode system with electrode gap sizes down to 20 nm and electrode thickness of 25 nm. Gold has a reliable surface chemistry, which is advantageous for antifouling protocols that utilize thiol chemistry, that can hopefully be applied in the future when using a surfactant is not possible. Gold also makes the electrode fabrication easier.

We have simulated the electric field in the designed gold nanogaps using COMSOL assuming a linear dielectric response to the field in a particle-free electrode structure and with de-ionized water as the medium (See Supplementary Information S1 for more information about the simulations). It is clearly visible that the field is concentrated around the tips of the electrodes. Figure 2A-B features the in-plane and cross-sectional distribution of the electric field. According to these simplified COMSOL simulations, fields can easily exceed $10^7$ V/m when applying only 1 V over a gap between two electrode tips that are 20 to 50 nm apart, which will keep the heating-induced temperature change well below 1 °C and thermophoretic effects negligible (Supplementary Information S2).

Figure 2C shows a scanning electron micrograph (SEM) of one gold electrode pair with a gap size of 28 nm, fabricated on glass following the protocol described in the Supplementary Information S3. Our chips were fabricated in batches of 4 on a large wafer. Each single chip (of 4 per wafer) includes 9 electrode pairs with gaps designed to range from around 20 to 100 nm. In our experiment we succeeded in trapping polystyrene nanoparticles from a batch of 20 nm average diameter particles (Figure 2E for the particle size distribution) while applying less than 1.5 V AC-voltage amplitude over the 20–50 nm gaps. We infer these gap sizes from SEM measurements of a small number of the samples (about 10) before applying any voltage. Gaps cannot be used for DEP after a SEM measurement because of carbon deposition, and DEP measurements themselves can modify the gap size, so that SEM measurements after DEP are not reliable either. We imaged one sample per batch right after fabrication, only as an estimate of the gap size in that batch. The electrodes used in the actual measurement on the batch of nominal 20 nm particles were unfortunately not imaged in the SEM afterwards. Therefore the electrode gap size is not precisely known but only within the variability between the chips within one batch, which we expect to be in the range of 20 nm. Based on this argument we expect the measurements to be performed on the gap sizes in the range of 20-50 nm .

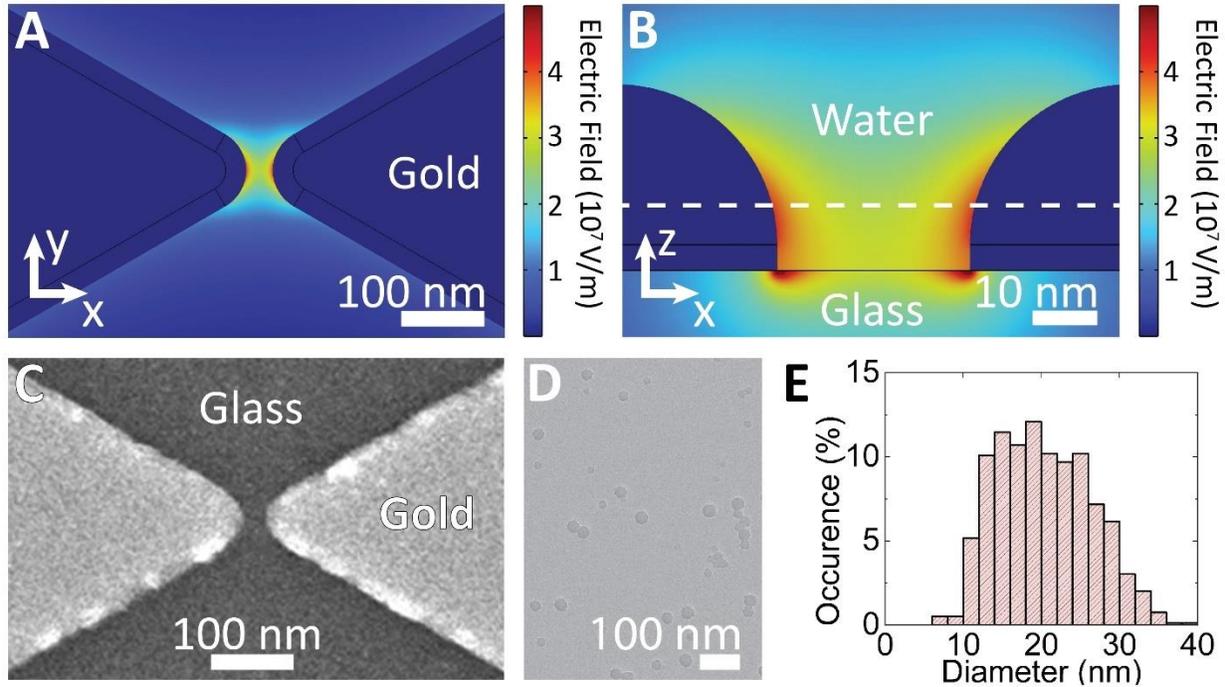

Figure 2. (A) Map of electric-field modulus in the plane of the nanoelectrodes simulated in COMSOL Multiphysics assuming the electrodes are immerged in de-ionized water. (B) Cross-sectional view of the electric field from the simulation in (A) where the white dashed line indicates the cut shown in (A) and is 10 nm above the glass surface. (C) SEM image of one of the fabricated nanogap electrodes. (D) Transmission electron micrograph (TEM) of our nominal 20 nm polystyrene particle batch. (E) Histogram of diameters of particles in the TEM image which has an average of around 20 nm.

To characterize our nano-DEP trap we used polystyrene fluorospheres from the average 20 nm batch (Figure 2D and 2E) as they are reasonably well understood from the viewpoint of DEP manipulation[53] and give both fluorescent and scattering signals that we can detect. The trapping potential for a polystyrene particle (Equation 2) depends on the frequency of the AC-voltage (Equation 3) mainly through the conductivity component of the complex permittivity. The conductivity of a polystyrene bead, $\sigma_p$, primarily arises from its surface conductivity[53]. In our experiments we used a water solution with a low concentration of a surfactant (See supplementary Section S12). We measured the solution to have a conductivity of 30 μS/cm. In environments with such low buffer conductivity, the surface conductivity arises mostly from the Stern layer[57] which is one part of the more complex EDL. The EDL dynamics involves many complex non-linear effects, which are still not completely understood for nano-scale particles and may significantly influence DEP[34]. We don't discuss these EDL effects in full depth in this paper but merely show a proof of principle of nano-DEP. As we increase the frequency of the AC-voltage applied across the gap, the real part of the effective polarizability (Equation 3) decreases. It eventually crosses zero at a specific frequency, known as the crossover frequency $f_X$, and then becomes negative, which causes the nanoparticles to be repelled from the point of highest electric field. For 44 nm latex nanospheres, $f_X$ is predicted to lie in the range 10-20 MHz[57,58]. As trapping of smaller dielectric nanoparticles using DEP hasn't been reported previously, we have trapped single particles at different AC frequencies to establish the crossover frequency for our nominal 20 nm polystyrene nanoparticle batch in our nano-DEP geometry.

Our initial step was to estimate the concentration of the nanoparticles which was low enough to allow for registering single particles in our nanogaps. In the experiment that will be described now we were observing fluorescence time traces with an AC-voltage amplitude of 0.75 V and frequency of 300 kHz for different dilutions of the stock solution of the average 20 nm diameter batch (Supporting information Section S7). The distinction between events from clusters and from single particles is visible in the fluorescent time trace, where the steps indicate the successive trapping of a second or more nanoparticles. By diluting the stock solution to a concentration of around $10^8$ particles/mL (which is around picomolar), we created conditions where single particles were trapped for several seconds without any new particles entering the trapping area. The following experiments were conducted at this concentration of particles.

During the first experiment we recorded time traces of the fluorescence signal in the nanogap while applying a range of carrier frequencies from 10 MHz to 100 kHz at an AC-voltage amplitude of 0.75 V. For each frequency, we maintained the conditions long enough to detect around 100 events, which were counted by the software by real-time thresholding of the fluorescence signal. The threshold was chosen to be 6.5 counts per 0.5 ms bin, which is 13 kcounts/s. If the particle resided in the trap for at least 1.5 s, the software automatically released it by setting the voltage applied over the gap to zero. We released particles after a couple of seconds because we knew from experience that the chance of clustering and irreversible sticking increased over time. Representative fluorescence time traces (Figure 3A) show a clear effect of the carrier frequency on the DEP trapping efficiency through the rate of occurrence of peaks and the dwell time of the trapped nanoparticles in the detection volume (Figure 3B). At 10 MHz, only weak fluorescent peaks are visible, but as the frequency decreases, the peaks become more frequent, more intense and extended in time. When the frequency was set to 600 kHz or lower, the duration of many trapping events extended beyond 100 ms, with some events lasting several seconds when the carrier frequency was under 600 kHz (see figure 3B). We noticed that events with a dwell time exceeding 200 ms also showed higher fluorescence intensity levels. This suggests that the longer trapping times might be associated with a subset of larger particles or that these particles reach closer to the gap area, which could increase the detection efficiency. We have extracted and compared the dwell times for the particles in the trap at different AC frequencies (Figure 3B) by looking at the boxplots fitting the data. Specifically, these boxplots cover the data in four parts that cover 25 % of the events each. Some outlier events are not in the boxplot range and are defined as events with a dwell time that is 1.5 times the interquartile range. At the 5 MHz and 10 MHz frequencies only events shorter than 35 ms were observed (the 0.5 ms binning of the fluorescence data prevents a more accurate estimation of these times). As these events are also similar in dwell time to the shortest events at lower AC frequencies, we can conclude that our confocal fluorescent detection volume is larger than the localization volume of the DEP trapping. While comparing these observations with the literature[57], we conclude that the crossover frequency for our particles lies between 1 MHz and 5 MHz, which is on the lower side of the literature prediction. However, the literature of polystyrene trapping is based on studies that have been conducted in much less localized field gradients and therefore a more linear response. Also those earlier studies were done at much higher particle concentrations than ours. Therefore, we use this previous work merely as a reference[57]. We expect that the events registered at the region of the negative DEP (5 and 10 MHz) are transient observations of particles traversing the detection volume. These diffusion times can be used as a measure for diffusive transport through the detection volume. For the rest of the paper we have been operating in the positive DEP regime: At 300 kHz for 20 nm average diameter and 100 kHz for 35 nm average diameter particles.

We note that in the dipole approximation it is assumed that the dipole is small compared to the length scale of the nonuniformity of the applied electric field[50] assuming the response of the medium and of the particle to be linear in the field, which may be incorrect in the high fields close to the electrode tips. In our nano-DEP experiments the particle and applied field are of similar lengths and higher multipole orders will therefore not be negligible. For spherical particles analytical solutions exist for an arbitrary higher order moment[54] and some corresponding multipole potentials can be found in work by T. B. Jones[50]. In analogy, the effect of these higher orders can enhance the trapping potential by a factor of about two in optical nearfield trapping at a tip, as stated by Hecht and Novotny in their textbook[59].

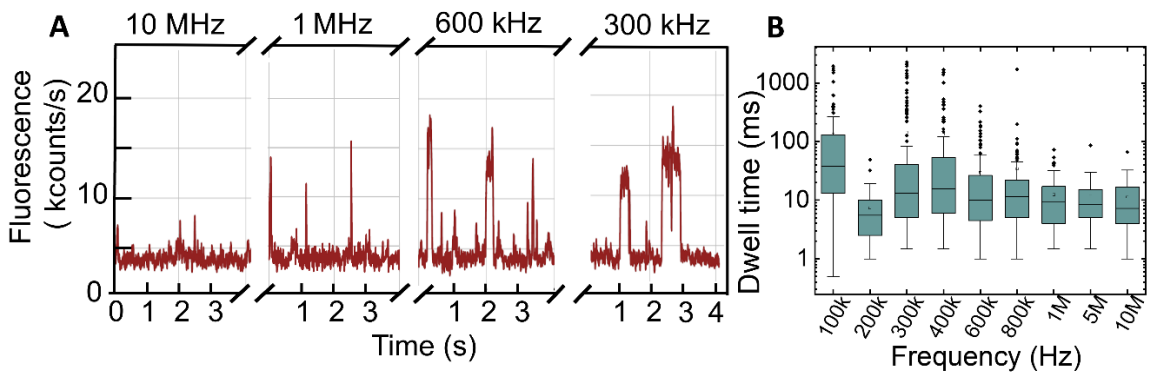

Figure 3 A) Fluorescence time traces for four different nano-DEP carrier frequencies during one experiment that started with 10 MHz and ended with 100 kHz carrier frequency. (B) Boxplots of the dwell time of single particle trapping events at all carrier frequencies we applied with a AC-voltage amplitude of 0.75 V. The black dots outside the boxplots are events with a long dwell time which we assign to genuine trapping events. At 200 kHz the behavior changes dramatically and the same happened again at 100 kHz. We attribute this change to damaging of the gold electrodes during the experiment time.

In the next experiments with a new batch of gaps we recorded both fluorescent and scattering traces from the nanogaps while applying the AC-voltage. Figure 4A shows a representative trapping event when trapping a particle from the 20 nm batch. The AC-voltage amplitude required was 1.5 V (see Figure 4A) which suggests that these gaps were closer to 50 nm than to 20 nm. The scattering data distinctly show when particles arrive at and leave the trapping area. These events happen synchronously with changes observed in the fluorescence signals. The difference in particle sizes within a batch –average 20 nm in figure 4 – was reflected in the average fluorescence intensity during one trapping event (12 kcounts/s to 25 kcounts/s) and the scattering intensity increase upon arrival (Figure 4B) which follows a linear correlation which fits the volume dependence of both types of signal.

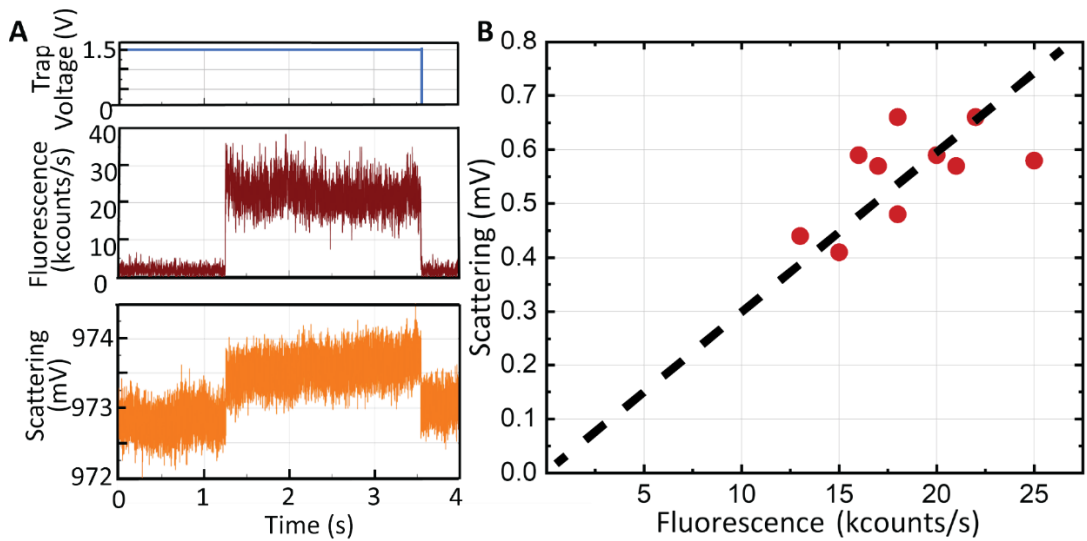

Figure 4 (A) Scattering (bottom), Fluorescence (middle) and waveform generator AC-voltage amplitude output (top) during a trapping event from the particle batch of 20 nm averagel diameter particles, trapped at 100 kHz. (B) Scatter plot of the steep scattering signal increase (y-axis) when the particle arrives (See bottom of panel A), against the average fluorescence intensity over a whole trapping event (x-axis).

Next, to double check if these are single particles we do a calculation on the range of iSCAT signal[12] changes that we can roughly expect from a single particle arriving in the trap (Figure 4B on the y-axis). First we calculate the expected signal (see Supplementary Section S9 for details of the calculation) for a particle of 34 nm in diameter, which we expect to be on the larger side of the size distribution of the 20 nm batch (Figure 2E) and therefore also expect to give the largest possible signal increase for particles in that batch. The metal layer of the electrodes shifts the phase of the reference field with an additional value. This value we add to the Gouy phase shift[1] of minus $\pi/2$ rad, that is accumulated by a focused Gaussian beam. We now use the interference term in Equation (1) to calculate the signal increase when a particle gets trapped. The interference term is $2|E_{scat}||E_{ref}|\cos(\varphi_{ref} - \varphi_{scat})$, and gives an intensity that is much larger than the pure scattering intensity from the particle ($|E_{scat}|^2$). The phase shift of the reference field turns the $\cos(\varphi_{ref} - \varphi_{scat})$ of Equation (1) into a value close to $\sin(\varphi_{scat})$. Both the experimental results in Figure 4 and the crude calculations of the phase shift that the reflected light undergoes at the metal layer, support our conclusion about the phase shift (when assuming we are focusing on the surface of the substrate). For a 34 nm particle, we assume $\varphi_{scat}$ to be a small positive value, given that we expect the center of the trapped particle to reside slightly above the glass (at least at between half of the particle diameter (17 nm) and the thickness of the electrodes (29 nm). Here we have taken, $\varphi_{scat} = \pi/4$, which is just an relatively small value that fits relatively well to our measured results. Considering the above argument on the phase shifts, the calculated/predicted signal increase for when this particle arrives in the trap is around $(1.5 \times 10^{-3})I_{ref}$. In figure 4A we can see that $I_{ref} = 973\ mV$ on the detector. The increase in signal due to a 34 nm particle arriving in the trap would then be equivalent to an increase of intensity by around 1.45 mV on the detector. In Figure 4B the signal increases are around half as big as calculated confirming that such scattering signals can be obtained from trapping single nanoparticles from 20-34 nm. It should be noted that the scattering contrast still has room for improvement by better phase matching between $E_{ref}$ and $E_{scat}$ (See Supplementary Section S10 for how the coverslip thickness can influence the detection efficiency for example). In the future, the model can be updated by performing more extensive simulations on the optical signals besides the presented back-of-envelope calculations that we did here. Looking ahead, we intend to fine-tune the plasmonic characteristics of our nanogaps and tune the laser to match the wing of the plasmon for each gap, which is not the case at the moment. Our aim is to improve the signal-to-noise ratio, ultimately enhancing our system's sensitivity to detect individual protein molecules without requiring labels.

Label-free protein detection would require a SNR 10-100 times greater compared to the scattering-signal-increase values shown in Figure 4B, as interesting protein sizes can lie in sub-10 nm range. Beyond standard methods of enhancing stability and sensitivity, our setup's ability to trap and actuate nanoparticles presents significant opportunities for SNR improvement. Highly-sensitive methodologies like photothermal microscopy[18], electric double-layer modulation microscopy[9] and super-resolution imaging[60] including iSCAT microscopy[12] use modulation of the useful optical signal to isolate it from the unmodulated background and to increase the SNR. With our setup, we can modulate the optical signal by slightly adjusting the position of trapped particles in the trap within the localized scattering detection volume, in contrast to the broader fluorescence detection volume (Figure 1C-E). We can impart a periodic motion to the trapped particle within the trapping volume through a periodic modulation of the trapping potential. We achieved this by modulating the amplitude of the AC-voltage of the carrier-frequency applied to the nanogap, using a carrier frequency value as in previous experiments (Figure 5A). In other words, the trapping potential of our nano-DEP trap varies periodically with respect to the level of the thermal energy, which periodically slightly loosens the particle's binding to the bottom of the potential (Figure 5A, potential). If the Brownian motion is not allowed long enough for the particle to escape the trap, it will be trapped again when the AM-modulated potential will be re-established in the next alternance. We have performed such experiments at 100 kHz and 300 kHz carrier frequencies, for various AM frequencies (the modulation depth was 100 %) ranging from 400 Hz to 1000 Hz (See Supporting Information Section S11 for all events from this experiment).

To start a trapping event, a particle has to enter the trap at the same time as the trapping potential exceeds the thermal threshold (Figure 5A). Because of this lowered trapping probability, less trapping events will be started and the chance of losing a particle that is weakly interacting with the potential will also reduce the chance of a stable trapping event (figure 4 is regular trapping). Yet during the time of an experiment, we still observed trapping and could analyze the behavior of the trapped particles in the modulated potential. Figure 5B features a fluorescent time trace of the most successful trapping and modulation event for a nanoparticle from the nominal 35-nm average diameter polystyrene particle batch (See Supplementary information S5 for TEM images). The carrier frequency applied was 100 kHz and the frequency of the amplitude modulation was set at 500 Hz for this specific event. The Fourier transform spectrum of the scattering signal when the particle resided in the gap (as confirmed by the fluorescence time trace) displays peaks at the frequency of the amplitude modulation and at its higher harmonics (Figure 5C). The presence of the higher harmonics can be attributed to the non-linear nature of the actuation in the highly non-homogenous trapping field (Figure 5A top) and to the nonlinearity of the scattering signal as a function of the position of the particle in the gap region. Moreover, the power spectral density of the scattering signal is consistently enhanced in the frequency range between 100 Hz and 5 kHz compared to the spectrum of the empty trap (Figure 5C). This broad extra noise spectrum can be attributed to the enhanced thermal motion in the trap caused by the periodic reduction of the trapping potential. The shape of this noise spectrum is well known in the case of a harmonic potential[61], such as an optical trap, but we are not aware of any analytical expression of this noise spectrum for our DEP trapping potential. Our trapping potential is not harmonic because eq. (2) doesn't easily relate the trapping potential to the particle's position. Indeed, the relation between the field squared and the position is complex (See figure 2), and the interaction with the surfaces is not taken into account. The prominent harmonics shown in Figure 5C suggest that line filtering could notably enhance the SNR as in a lock-in experiment. Figure 5D compares the raw scattering time trace with traces filtered through a 1 kHz low-pass filter (in red, matching the fluorescence trace resolution) and a 500 Hz line-filter (in purple). The SNR of the line-filtered iSCAT trace is at least twice as high as that of the low-pass-filtered trace and 5 times greater than that of the fluorescence trace of a similar trapping event without modulation (Figure 4B). Such drastic improvement in SNR shows promise for downscaling our system and moving onto trapping of single protein molecules.

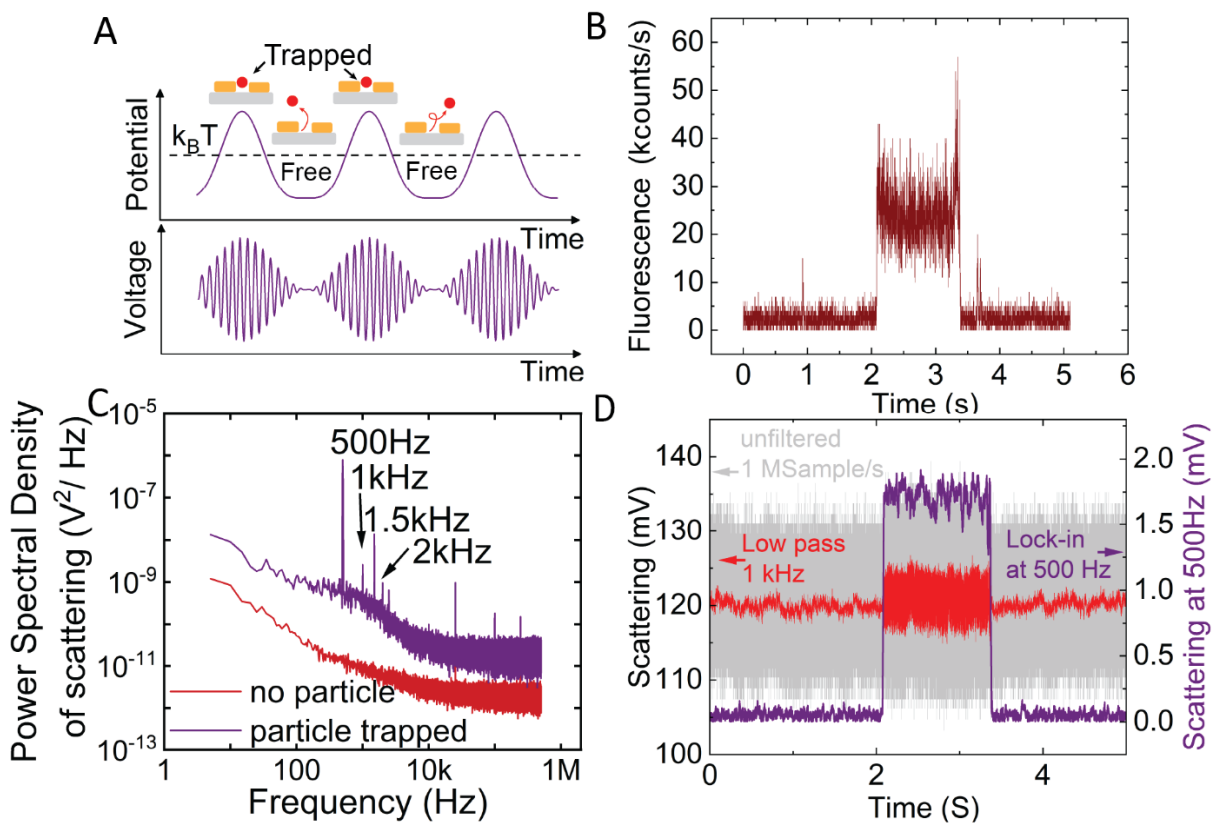

Figure 5 (A) Schematic layout of the experiment. Fluorescent polystyrene beads from the 35 nm diameter average batch are trapped and actuated in the gap. (B) Fluorescence time trace of the trapped particle in the modulation regime as in (A). (C) Fourier spectra of the scattering signal

with (purple) and without (red) the trapped particle in the trap confirming cross-talk of the modulation frequency into the optical signal. Note that for the sake of visibility we have shifted down the red values by a factor of 10. (D) Time trace of the scattering signal from the nanogap with different filter settings: raw trace at 1 MHz sampling (1 MSample/s) of the detected signal (light gray), low-pass-filtered time trace (Butterworth filter of 6th order) at 1000 Hz (red), and line filtered with a digital lock-in amplifier at 500 Hz (purple).

## Conclusion

To conclude, we have introduced a new tether-free strategy to prolong the observation time of analytes in confocal iSCAT microscopy. We have fabricated gold nanogaps to control the position of the nanoscopic analytes by means of nano-DEP. The nanogap electrodes allowed us to trap single polystyrene beads down to 20 nm in nominal diameter with AC-voltages amplitudes not exceeding 1.5 V, thereby avoiding heating and chemical reactions at the electrodes. DEP trapping and release enables several seconds-long observation times in both fluorescence and scattering. Finally, we have shown that periodic modulation of the trapping potential can impart a periodic change in the motion of the trapped analyte particles, which enables lock-in detection and provides an enhanced signal-to-noise ratio. In the future the improved detection will be applied to the study of smaller objects in view of single-protein trapping and detection. Considering recent progress in the understanding of protein DEP[62], our methodology represents a promising tool for characterization of proteins through their DEP response, which goes beyond the one dimensional mass sizing by iSCAT. In the future, our tool could open up avenues into the investigation of protein dipole moments and complement the iSCAT signal with structural and charge parameters of trapped proteins or other nano-scale analytes.

# Methods

The combined nano-DEP and confocal iSCAT method required a merged setup consisting of an electronic input connected to an AC-voltage source for dielectrophoresis and a home-built inverted confocal microscope enabling both interferometric scattering and fluorescence detection. Nano-gap electrodes were fabricated on a glass coverslip. This coverslip was mounted on a heavy steel disk with a hole in the bottom to point the microscope objective through. There the electrodes could be connected to a custom made printed circuit board which contained switches for short-cutting the traps when they were not in use (See Supplementary Section S3 for more details about the fabrication and the printed circuit board).

**Nano-electrodes and electronic design**

Briefly, the fabrication of the nano-electrodes is carried out on an approximately 200 μm-thick wafer. The electrode pattern, comprising of an approximately 4 nm thick chromium and 25 nm gold layer, is achieved through UV lithography for the larger structures and e-beam lithography for the tips of the nano-gap (See Supplementary section S3 for details). Each sample contains nine gaps formed by the tips of two bowtie shapes that have a space between them of 20 to 100 nm, depending on the purpose of the experiment. Each side of the electrodes ends with a large gold patch (~2 mm diameter) to ensure the connection to the printed circuit board (PCB), which manages the voltage applied to the electrodes through a system of switches and thereby protects the nanogaps from static discharges.

The generation of the AC-voltages is performed by two different function generators. For the data in Figure 1 to 4 it is a USB multifunction instrument (Analog Discovery 2, DIGILENT) with a sample rate of 100 MHz, 1 MΩ input impedance, and a ± 5 V voltage range. This device has the advantage of changing the voltage or frequency without interrupting the waveform. For the signal detection we used our NI-9775 digitizer module (a module for compactRIO of Labview, National Instruments) in high resolution mode settings for the data in Figures 1 to 4. For the data in Figure 5 a Siglent SDG1032X waveform generator was used to generate amplitude modulated signal and we used the same DAC but in "high speed" mode. This mode presents additional 1/f noise into the readout of scattering, which was useful for demonstrating the purpose of modulating the signal in a noisy environment, as can be seen in figure 5.

**Optical setup**

The optical setup is based on an inverted confocal microscope with a HeNe laser ($\lambda$=633 nm) as the light source. The light is focused through an oil objective (Olympus 60x, NA 1.4) fixed on a steel plate. *xyz* translation is performed by a piezo-stage with a range of 80 μm on x and y, and 20 μm on z. The ensemble of the objective, stage, and sample is supported by a rigid four-pillar cage pressed against a layer of rubber to damp vibrations. Both scattered and fluorescence emission is collected by the same objective and separated by a dichroic mirror (DM) (LP>650 nm). The reflected light (i.e., scattering) is focused through a 50 μm pinhole followed by a photodetector (Femto photo diode, Si-Pin). The transmitted beam is filtered with a set of bandpass filters (653-712 nm) and a 75 μm pinhole before it reaches a single-photon counting module (SPCM, Perkins-Elmer). (See Supplementary Section 13 for the more detailed optical setup).

**Polystyrene nanoparticles**

In this study, we used polystyrene spheres with nominal diameters by the manufacturer of 40 nm (F8789, ThermoFisher SCIENTIFIC) and nominal diameter by the manufacturer of 20 nm (F8783, ThermoFisher SCIENTIFIC). These beads are surface-modified with carboxyl groups resulting in a negative surface charge. Additionally, both sets of beads incorporate fluorescent dyes exhibiting maximum absorption/emission at 660/680 nm, to enable detection of these particles through the fluorescent channel of our microscope. To prevent aggregates in the experimental process, the beads are diluted in MilliQ water and subjected to a 5-minute sonication. The Transmission Electron Microscopy (TEM) images of the nominal 40 nm beads reveal a broad size distribution and a sifted mean diameter, measured at 33 ± 13 nm. Conversely, the 20 nm beads closely adhere to specifications, exhibiting an actual size of 21 ± 6 nm. The distributions present an almost equal probability of encountering beads from 13 to 25 nm in the nominal 20 nm stock solution.

**Trapping and releasing of single nanoparticles**

Manipulating nanoparticles with precision demands real-time control over processes occurring at millisecond timescales, which is beyond the scope of manual intervention. To achieve the trapping and releasing of individual nanoparticles, a LabVIEW-programmed sequence of actions is orchestrated, providing a repeatable procedure. Three measurement channels are concurrently recorded to investigate events: fluorescence, scattering, and the waveform generator signal. The fluorescence channel acts as a probe to detect the presence of a nanoparticle in the trap. Exceeding an intensity threshold triggers a certain pre-programmed sequence of actions by the waveform generator that applies voltages that are over the gap between the electrode tips. The scattering channel functions as a readout, enabling the characterization of label-free measurements. The waveform generator channel is employed to regulate the AC-voltage applied to the electrodes. A range of input parameters (peak-to-peak voltage applied to the electrodes, AC frequency, fluorescence intensity threshold, and trigger/release duration) must be optimized and calibrated to perform nanoparticle trapping, detection and characterization.

**DEP Spectroscopy and actuation methods**

To comprehensively understand nanoparticle behavior within an AC field, we explored different trapping voltages and frequencies using polystyrene particles of 33 and 20 nm average diameters (From inspection by TEM). These particles are suspended in a solution of MilliQ water and 0.5% nonionic surfactant (Tween 20, See Supplementary section S12) to prevent non-specific sticking to the electrode surface. This ensures escape of the particles from the trap when the AC voltage is set to zero and allows studying hundreds of particles at one trap within one single experimental run. In the context of non-biological samples, such as polystyrene nanoparticles, the introduction of a nonionic surfactant proves effective in preventing any adhesive interactions. Using these electrodes with proteins would require using antifouling coatings on the gold surface instead of surfactants[63]. To avoid simultaneous trapping of multiple particles, the solution concentration is maintained at approximately $10^8$ particles/mL. The solution was placed on top of the electrodes and was contained by a 200 μL open PDMS chamber which was sealed with a removable lid such as a small coverslip, that prevents leaks but allows an easy removal for the cleaning procedure of the sample between experiments.

The cleaning procedure that we performed after sample fabrication, before we did nano-DEP experiments consisted of a so called RCA-1 cleaning procedure. This is a procedure that is used to clean chips from organic material and other small particles. The samples were placed in a mixture with volume parts (5:1:1) of MilliQ:Hydrogen-peroxide:Ammonia-water respectively. This was heated to 75 °C and left in this solution for 30 to 60 minutes, while the fluid was slowly stirred by a magnetic rot. After an experiment the shortcuts would not be on the sample anymore so electric discharge was a concern during cleaning of the samples between two experiments. To prevent discharge we used a wristband that was connected to a high impedance and also to the surface that we placed the cleaning equipment on top of. This prevented discharges when treating the sample carefully during the cleaning. After cleaning and then drying of the sample in a nitrogen stream the samples were remounted on the PCB where they could again be shorted till an experiment would be started. The samples did not discharge when they we simply covered in water and no high voltage source, such as a human, was touching it.

To avoid restraining the particle and to be able to do a digital lock-in detection, a weak trapping is possible through amplitude modulation of the voltage applied to the gap, as described with the following equations:

$$S(t) = A_c[1 + k \times m(t)]cos\,(2\pi f_c t) \qquad (4)$$

$$m(t) = A_m\,cos(2\pi f_m t) \qquad (5)$$

Where $A_c$ is the amplitude of the carrier oscillation at a frequency $f_c$. $|k.m(t)|$ represents the depth of the amplitude modulation $(1 \leq k \leq 0)$, $A_m$ and $f_m$ represent the amplitude of the modulation and its frequency. When the potential fluctuates between two well-tuned values, the particle periodically moves away and returns to the gap if the amplitude modulation frequency is higher than the escape rate. To examine the impact of weak trapping on particle behavior we refer to figure 5 and the Supplementary section S11.

AUTHOR INFORMATION

**Corresponding Author**

Sergii Pud

**Author Contributions**

The manuscript was written through contributions of all authors. All authors have given approval to the final version of the manuscript. ‡Jacco Ton and Théo Travers contributed equally.

ACKNOWLEDGMENT

SP, JS, DW acknowledge The Netherlands Organisation for Scientific Research (NWO/OCW) for financial support in a form of VENI grant 2019 (VI.VENI.192.162) and ENW-XS-23-4 (OCENW.XS23.4.129), MO thanks the Spinoza Prize (2017) for financial support. The authors thank the Kavli Nanolab (made available through the NanoFront Programme) and MESA+ Nanolab. Authors thank Eliot Schwander, Lucas van der Togt, and Janike Bolter for valuable discussions and input into development of the project. Generative AI software (ChatGPT Plus, OpenAI) was used to assist with generating code for data visualization. The output of the generated scripts was validated and debugged.

## Supplementary information

**Section S1 COMSOL Simulations of the electric field in the nanogaps**

We used the COMSOL Multiphysics finite element modeling to simulate the electric field generated by the electrodes in a minimal geometry of the experiment. The 3D model geometry includes two equal blocks of water and glass (5×5×1 µm), and a pair of equilateral triangle (side=500 nm) gold electrodes placed on the glass surface, separated by 20 to 50 nm gaps. We imported the simplified gold layer geometry made in Autodesk Fusion 360, following the rounding of the tips based on the SEM images of the fabricated samples. The gold layer thickness was chosen to be 29 nm, equal to the fabricated sample thickness including the Chromium layer. The edge curvature is considered due to the nature of the liftoff process, which tends to wound the upper edge of the metal film. We employed a custom triangle mesh with element sizes ranging from 0.1 nm to 200 nm. In the regions close to the sharp edges, we used a finer mesh, reducing the maximum mesh size to 1 nm, to get more precise results. To calculate the electric field modulus we used the stationary solver in the electric currents module. The DC voltage difference between the electrodes was set to 1 Volt. We used material properties as listed in the COMSOL Multiphysics library.

**Section S2 Heating of the electrodes**

The Joule heating originating from the conductance of the electrolyte and the double layer is a serious concern when designing a DEP[1] experiment. The heat flux from the electrodes is balanced by heat dissipation through thermal conductivity of the surrounding media and the substrate. Ramos et al.[2] provides an 3D analytical solution in spherical coordinates assuming a set of flat linear electrodes separated by a distance $l$. The solution yields the maximum temperature difference between the infinity and the point of the Joule heat generation:

$$\Delta T \approx \sigma \frac{U^2}{8k}, \tag{S1}$$

where $\sigma$ – is the electrical conductivity of the medium, $U$ – is the voltage applied and $k$ – is the thermal conductivity, which equals 0.60 W/(m×K) at room temperature for water. The electrical conductivity ($\sigma$) that we take as input was measured in our medium solution to be $3 \times 10^{-3}$ S/m (our device gave as output 30 µS/cm). At such low conductivities of the electrolyte, the surface conductance related to the ions screening the surface charge become a significant factor. We can estimate this conductance when we know the surface charge of the glass surface in the nanogap. According to Behrens and Grier[3] we can expect a surface charge density, $\rho_s$, of -0.2 mC/m² for a pH of 7 and ion concentration below 1 µM. These negative charges are screened in the solution with sodium ions that have electrophoretic mobility $\mu_{Na^+} = 5.2 * 10^{-8}$ m²/V/s. These ions can move in the response to the electric field along the glass surface and will yield additional conductivity $\sigma_s = \frac{\rho_s \mu_{Na^+}}{h_{el}}$, where $h_{el}$ is the height of the electrodes (29 nm). Plugging in the numbers yields 3.5 µS/cm, which is 10 times less than the conductivity measured for our solution. This number does not take into account the conductivity of the double layer around gold, but this should be of a similar order of magnitude as it is also slightly electronegative at neutral pH. Therefore we can proceed with our estimation using the measured conductivity of the solution. The AC-voltage amplitude applied between the electrodes is $U$ = 4 V here, which is more than the maximum voltage we applied in practice. Since we are working in water, we can estimate the thermal conductivity $k$ at room temperature to be 0.60 W/(m×K). The highest temperature change can be measured perpendicular to the surface and theoretically gives us a maximal heat change of $\Delta T \sim 10$ mK. This value can still be considered a lower bound estimate due to potentially unaccounted factors, but it is significantly low and considering that we have never observed any kind of bubble formation at AC voltages of 4V, we consider the heating effects of our electrodes negligible.

**Section S3 Fabrication of the gold nanoelectrodes**

As substrate a wafer was used (Schott, borosilicate glass, MEMpax, 4 inch diameter and 200 to 240 µm thickness) with surface dimensions that can fit six samples of dimensions 24 mm × 24 mm. We enhance the chance of obtaining several samples with comparable gap sizes within one batch by making and handling many samples on one substrate that goes through the whole fabrication procedure as one piece. The protocol utilized two different lithography methods during the fabrication, one for larger structures and one for the nano-scale tips. This led us to divide the clean-room work into two fabrication round trips for the substrate. Round one for UV-lithography of the larger structures (Contacts and leads shaped as in figure S1a-b) and round two for the e-beam lithography of the small tips with nanogaps between them. Figure S1b-g shows how the sample looks after the different steps of the e-beam fabrication round (which is the second round). Those drawings (Figure S1b-g) specifically depict the intended results of the fabrication steps for the small tips, but the UV-lithography round (the first round) for making the larger structures follows roughly the same procedure of steps, as explained before the explanation of the Ebeam-lithography round.

The first round (Large structures fabrication with UV-lithography):

1) After we had designed our structure in the software used by the lithography machines (Figure S1show the design) we could start the fabrication procedure. The first step (Substrate cleaning) was started by twice placing the wafer in a 99% $HNO_3$ solution for 5 min, to remove traces of organic material. Next the wafer was rinsed in DI water. The wafer was subsequently dried in a nitrogen stream. Now the substrate first underwent a dehydration bake (120 °C and 10 min) on a hotplate and then was placed on the spincoater (device: Primus SB15 Spinner), where it was liquid-primed with hexamethyldisilazane (HMDS) to improve the hydrophobicity of the glass (settings: static spin mode, 4000 rpm, 30 s).

2) A positive UV photo-resist (Olin OiR 907-17) was spin-coated on top of the substrate (settings: static spin mode, 4000 rpm, 30 s) to a thickness of 1.7 µm. Removal of residual solvent from the resist was ensured by a prebake at 95 °C for 90 s on a hot plate.

3) UV-lithography was done by using a mask placed centrally on the wafer plane with a mask alignment system (EVG®620 NT) which also contained a UV light source (settings: separation 50 µm, hard contact mode, constant time exposure mode, exposure time 4.2 s, Hg lamp power of 12 mW/cm$^2$).

4) Development of the Olin OiR 907-17 resist was done by dipping the wafer into two different beakers after each other, both filled with OPD4262 positive photoresist developer (30 s in the first beaker and 15-30 s in the second beaker). The sample was again rinsed with DI water. A post-bake was done on a hotplate (120 °C and 10 min).

5) A chromium (~4 nm thick) adhesion layer and a ~25 nm gold layer were deposited on the developed sample. Cr and Au were deposited by DC sputtering by two different sputtering machines (in house at MESA+ in Twente) machines. Before the Au was DC sputtered onto the sample a pre-sputtering was done for removal of surface contaminants from the target.

6) The fabricated pattern was cleared of unbound metal and residual resist by means of a lift-off process: The sample was placed in a beaker with acetone and placed in a sonicator for 2 min. Then it was left overnight in acetone to let the metal flakes precipitate. After lift-off the sample was dried in a nitrogen stream (see fig S1 for the designed results in cartoon form).

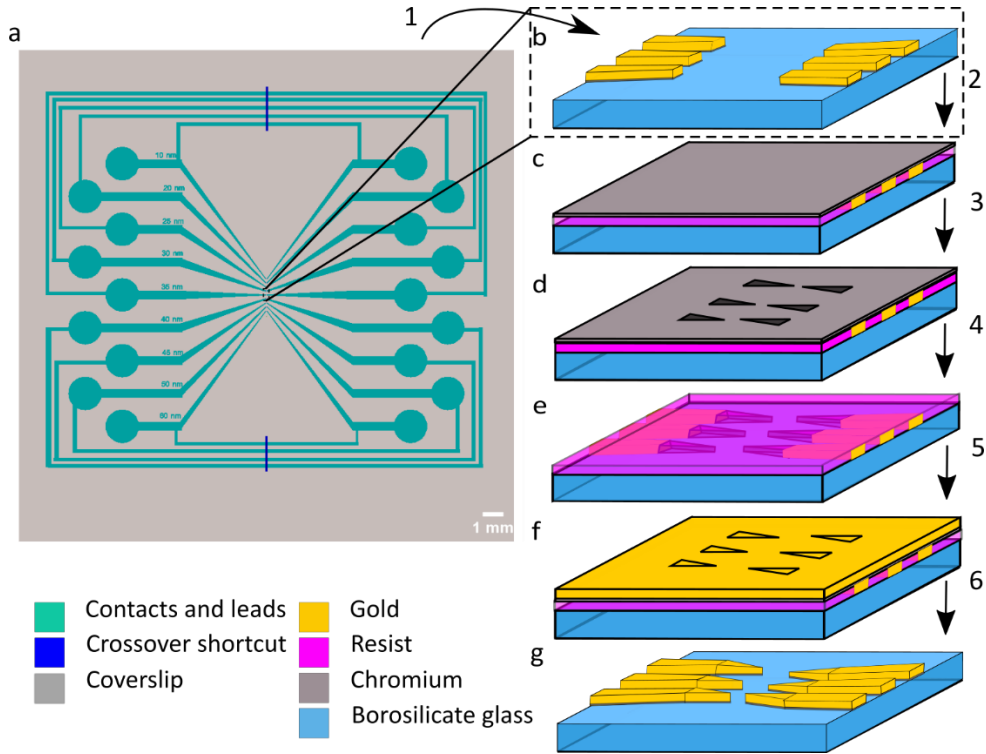

Figure S1: This figure shows the 6 steps of the UV-lithography fabrication process described below. (a and b) UV-lithography results for the large scale patterns. (c-g) The results after the steps taken during the e-beam lithography round of fabrication. The substrate is different and the e-beam round is done on a substrate that is partly covered in gold structures, which can influence the end results of the fabrication procedure especially around the connection between the large structures and the small structure from the Ebeam-lithography.

After fabrication of the larger structures by UV-lithography (contacts and leads), the sample was prepared for e-beam lithography of the nano-gap tips (which can only be seen when zooming extensively in on the design in figure S1a).

1) The wafer, plus larger structures, was dehydrated and primed with HMDS as for UV lithography.

2) After priming, the sample, with the large structures already fabricated, was spin-coated with a positive e-beam resist (CSAR AR-P 6200.04, at room temperature) of 160 nm thickness (settings: 2000 rpm and 45 s). Directly after this the sample underwent a soft-bake (155 °C for 2 min) for residual solvent removal. A ~8 nm thick chromium layer was DC-sputtered on top of the resist (see figure 1c).

3) A RAITH EBPG5150 machine was used for e-beam patterning (see figure S1d for e-beam patterns) of the resist (settings: beam limiting aperture 300 µm, current: 1.5 nA, dose: 200 µC/cm$^2$, with the BEAMER software set to Multipass exposure and a short range correction of 10 nm).

4) After e-beam exposure the Cr was etched away with chromium etchant. Then the pattern was developed in a beaker with pentylacetate for 60 sec followed by 60 sec in a beaker with methyl-isobutylketone (MIBK): isopropanol (1:3 volume%) and then the samples were rinsed for 30 s in a beaker with isopropanol. The samples were blow-dried afterwards in a nitrogen stream. A descum step was performed in UV-ozone (PRS100-ILP) for 300 s to remove residual resist in the developed areas (See figure S1e for the sample after development).

5) After the descum step, deposition of Cr and Au was done (see figure 2f for the result) in a Balzers BAK600 e-beam evaporation device. First a chromium adhesion layer was evaporated onto the substrate (5 nm thick, base pressure $<1.0\times10^6$ mbar, 8 kV beam voltage) and then the Au layer (25 nm thick, base pressure $<1.0\times10^6$ mbar, 10 kV beam voltage).

6) Lift-off was done in a beaker with acetone at 45 °C while leaving the sample there for a few hours and moving it up and down in the fluid a few times, without further stirring the fluid actively. The wafers were diced (Nitto SWT 10 dicing foil, DAD dicing saw or Load point Micro Ace 3) and rinsed in DI-water in a sonicator, to remove particles, after which the sample was dried, inspected with an optical microscope if deemed important and then packed in antistatic foil for storage and transport to the optics lab.

The printed circuit board, that we could connect to our connector patches of the electrodes, was connected through coaxial cables to a function generator. The design of the board was as shown in figure S2. The four holes in the corners of the red structure are to put small screws through to connect the sample to a sample holder which is a steel disk with a hole in the center.

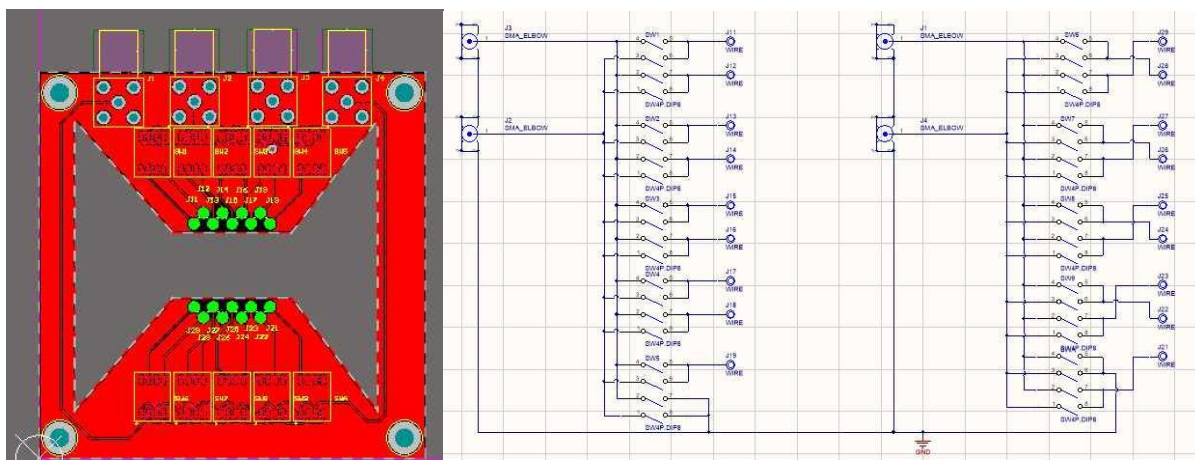

Figure S2: (Left) The printed circuit board seen from the top. The green spots are the spring connectors that will softly push down on the electrode connector patches (as designed in Figure S1a). (Right) the wire diagram of the board. We only used 2 but it has the possibility for 4 electrodes connected at the same time.

**Section S4 Electrode degradation after the RCA cleaning**

It should be noted that after each measurement session, the sample was cleaned using a part of the first part of the RCA cleaning procedure called SC-1, which is described in the methods section. This could cause degradation of the electrode and widening of the nanogaps until trapping at low voltages was no longer possible. SEM images of these nanostructures also show irregular degradation of the gold layer, giving it a mesh-like structure. This RCA etching procedure is very slow on gold but when the chemicals come in contact with organic material it will quickly dissolve the organic material. This can also cause bubbles and degradation of the electrode structure when the chemicals can reach under the structure and reach left-over e-beam resist there. Next a SEM image is presented (Figure S3) of an electrode structure that was used to trap relatively large particles of 33 nm or more. Because this gap is more than 100 nm wide we expect the experiments presented in the paper on our 20 nm particles and which were performed at lower AC-voltage amplitude to have been done with smaller gap sizes.

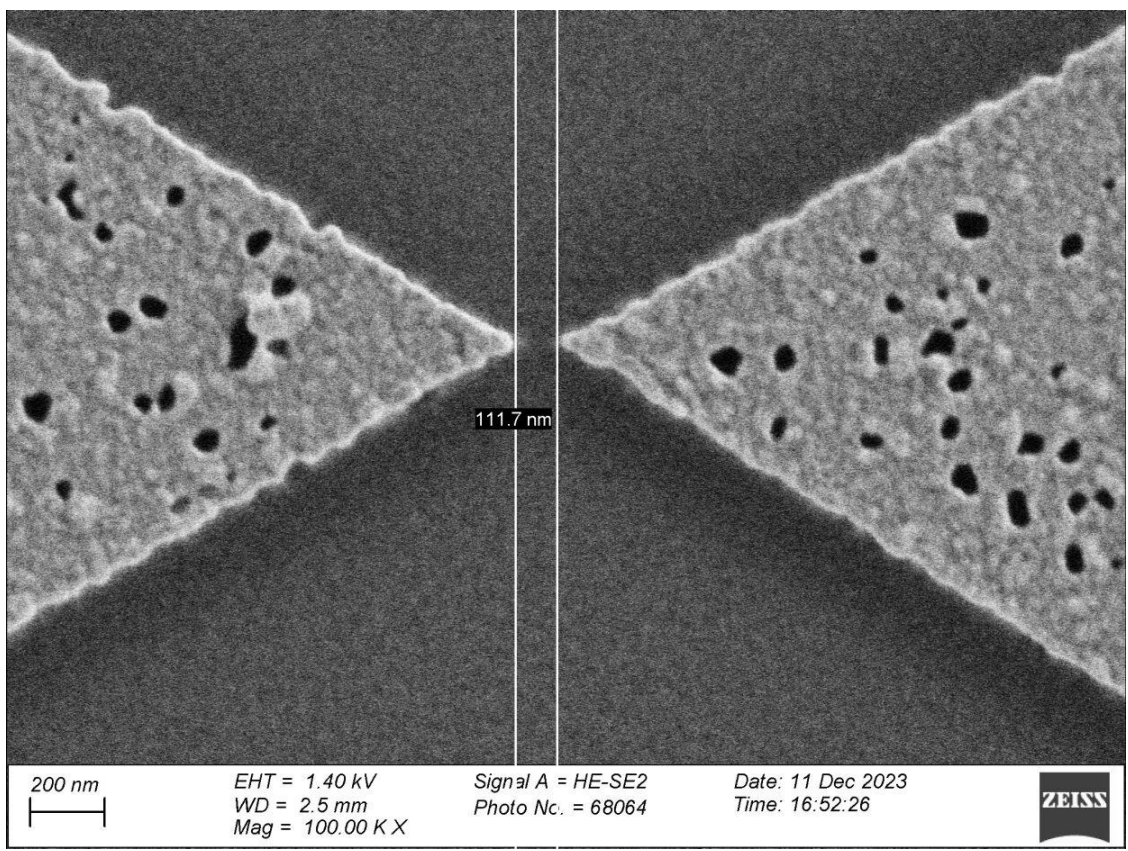

Figure S3: A SEM image (settings in the legend at the bottom) of a gap that was designed to be around 30 nm but which after the experiment was found to be as shown here.

**Section S5 TEM images of 20 and 33 nm average diameter polystyrene beads**

Transmission Electron Microscopy (TEM) images of the beads (we checked both stock solutions that we have which the manufacturer sold as nominal 40 and 20 nm in diameter) reveal a broad size distribution of $33 \pm 13$ nm and $21 \pm 6$ nm respectively (See figure S4). The 20 nm beads closely adhere to specifications, exhibiting an actual size close to 20 nm average. The distributions of the sizes are not gaussian and almost represent equal probabilities of encountering beads from 13 to 25 nm. We still use the mean from a Gaussian fit to extract what we here call our mean diameter. The 33 nm average diameter beads have a broader distribution.

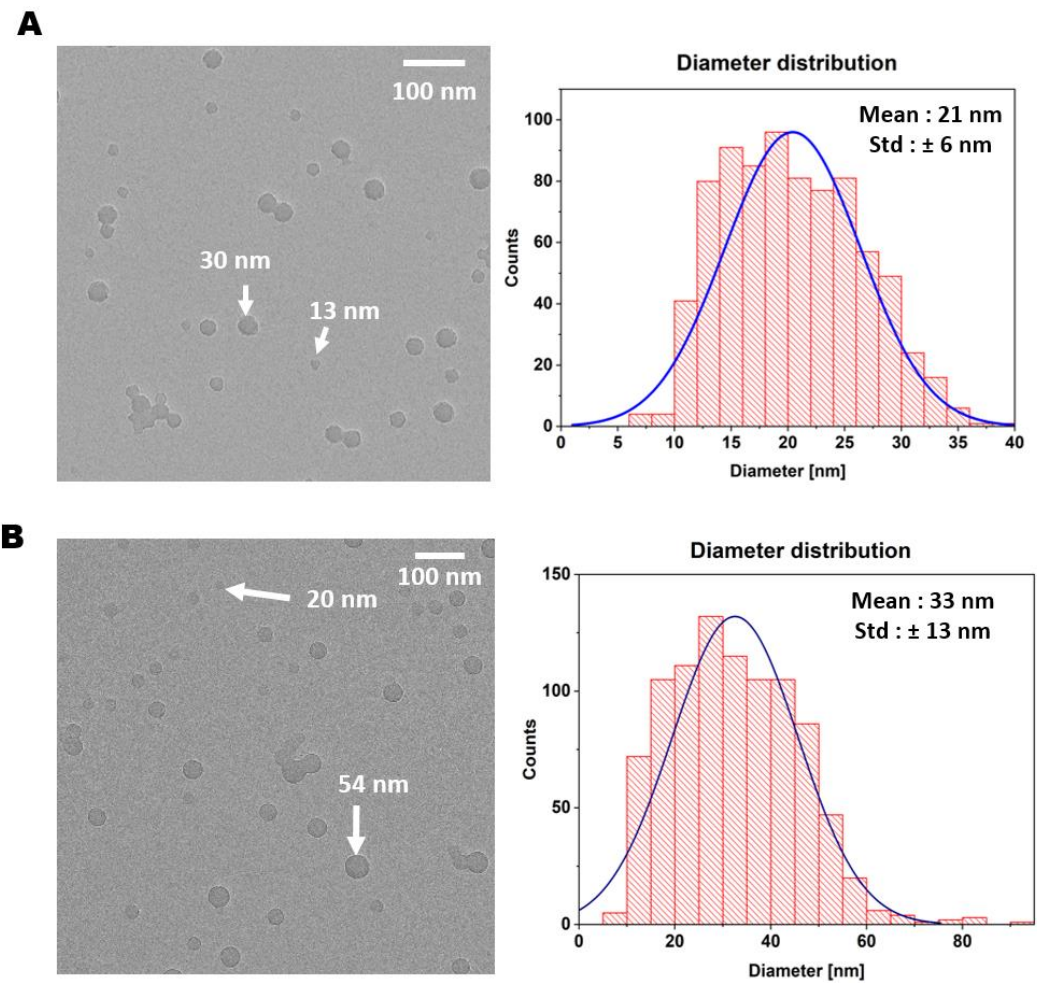

Figure S4: TEM images and size distributions of (A) 20 nm average diameter (according to the vendor also 20 nm) beads and (B) 33 nm averagediameter beads (40 nm according to the vendor).

**Section S6 Single-particle fluorescent counts and scattering signal**

To be reasonably convinced that our detected events are from single particles and not clusters, figure S5 shows 33 nm fluorescents beads (See Figure S4B TEM data for their size) that were spin-coated on a coverslip at 5000 rpm for a 60 s in a solution of distilled water mixed with 1% of polymethyl methacrylate (PMMA) to settle the particles on the coverslip before they cluster when the solution evaporates during the spin coating. This means that after spin-coating we have single particles to look at in the microscope

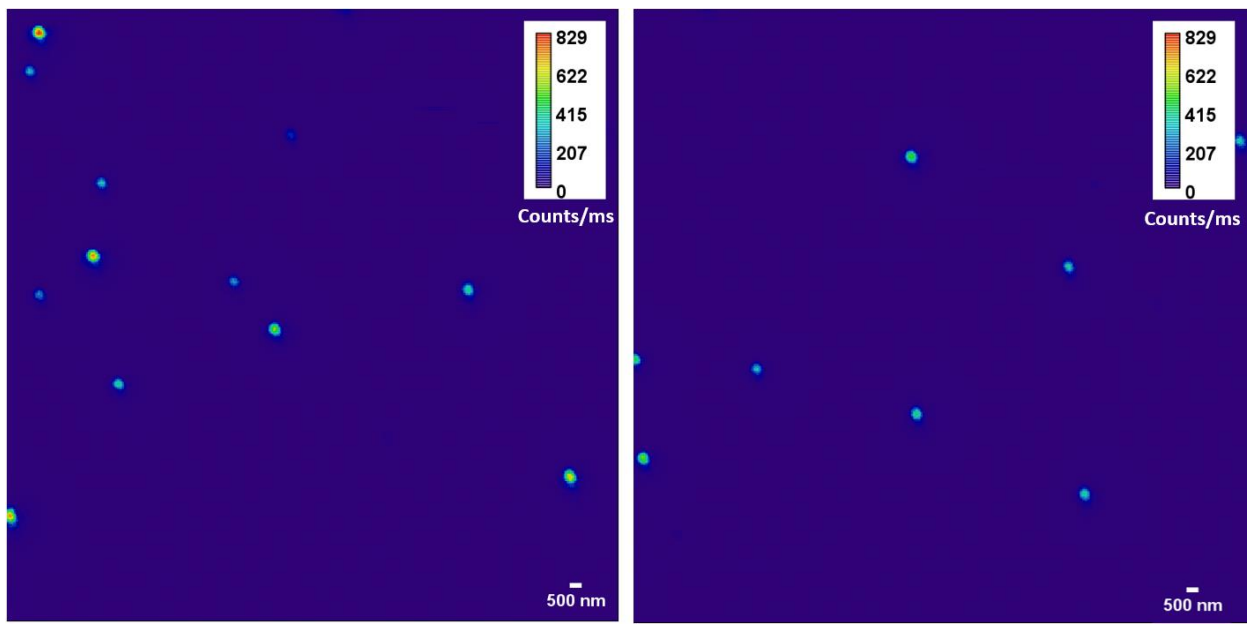

Figure S5: Image of an area of 25.4 × 25.4 µm² with fluorescent beads of 33 nm nominal diameter spin-coated on a standard 170 µm thick coverslip. Laser power 12 µW (which is close to the illumination used in our experiments in the main part of this paper) and 1 ms of integration time.

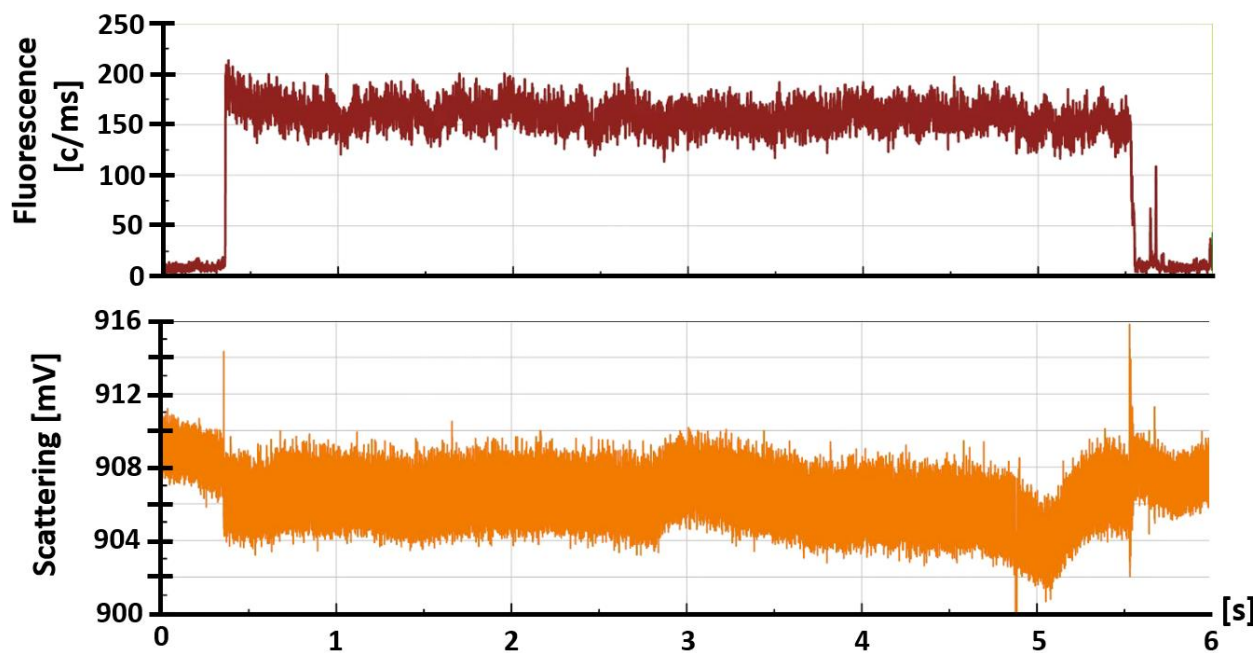

Figure S6: Fluorescence (top) and scattering (bottom) time traces of a 33 nm diameter particle. It is also interesting to see that this trapping event could last more than 4 s and that the interference here was negative. We believe that during this single experiment the focus position was different and therefore changed the phase shift, creating a destructive interference instead of the common positive interference.

Many fluorescent events from 33 nm polystyrene particles show counts in the trap of less than 150-300 counts/ms. Possible reasons are: i) shadowing effect because of the gold structures, ii) changes of the illumination PSF because of the thicker glass; we observed less than half the intensity when using thicker wafers of 200 to 240 μm thick. Another interesting effect is visible in figure S6 in the scattering time trace that goes down in intensity when the particle arrives or leaves. This can happen when we had a particularly large wafer thickness which made it hard to find the focus at the glass gold interface. This can be qualitatively be explained as follows. When the z position of the coverslip moves up or down in the focus the acquired Gouy phase of the reference field will change which can even flip the sign of the cosine term in the detected scattering intensity equations (Equation 1 in the main text).

**Section S7 Nanoparticle dilution series to identify conditions for single-particle trapping**

To perform a repeatable and reliable measurement of single particle trapping, an empirical approach is suitable due to the conditions that can change between two experiments (gap size, particle size, solution pH can all change between experiments). The fluorescence channel can help to determine the right concentration of particles to avoid accumulating them in the gap. The average time between two successive bursts in the fluorescent channel indicates if the concentration must be reduced for the time that we aim to trap for. As the left part of figure S7 shows, a too high concentration of particles leads to the accumulation of the particles in the gap with additionally a bigger risk of irreversible sticking of particles to the surface.

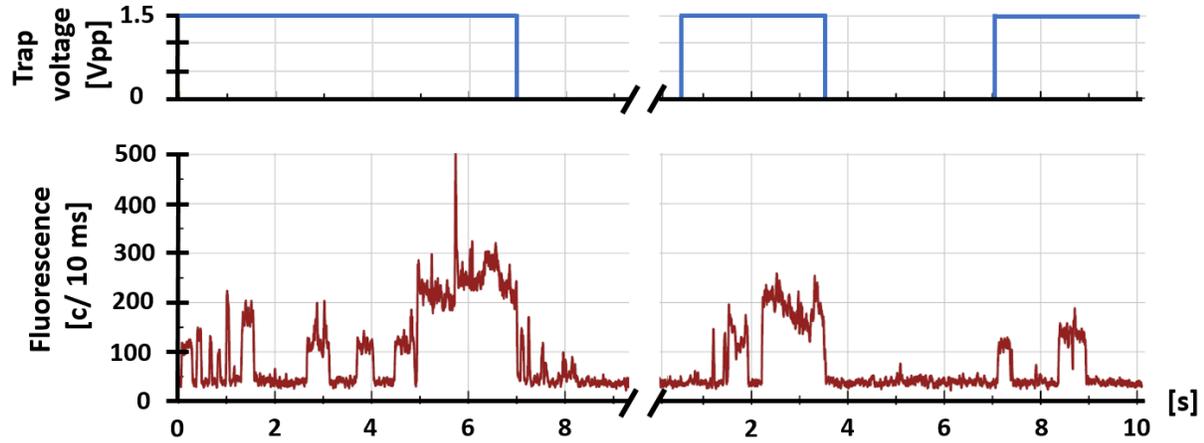

Figure S7: Fluorescence time trace indicating particle clusters due to too high a concentration. The second part is the same experiment with a factor 5 less particles in the solution, which enables single-particle trapping.

**Section S9 Nature of the scattering signals intensity change from the trapped polystyrene nanoparticles**

First we calculate the dark-field scattering cross section ($\sigma_{scat}$) and scattering coefficient ($s$) for a small scattering particle. The steps are, 1) calculate the volume of the particle ($V_p$), 2) calculate the particle's polarizability while assuming a small sphere in a plane-wave illumination ($\alpha_{Rayleigh}$), 3) calculate the scattering cross section, 4) divide the scattering cross section by the illuminated confocal spot area ($A$) that we illuminate during the experiment. The result will yield the scattering coefficient, $s^2$ which is the factor used to calculate the $|E_{scat}|^2 = s^2 I_{inc}$ with $I_{inc}$ - the intensity of the incident field of the sample illumination.

In equations this looks as follows, with the results for a particle of diameter 20 nm and a particle of 34 nm presented respectively[4,5].

The volume is, when using the particle diameter ($r$):

$$V_p = \frac{4}{3}\pi r^3 \tag{S2}$$

$= 0.42 * 10^{-23}\ m^3$ and $2.06 * 10^{-23}\ m^3$ respectively.

The polarizability is (with $n_s$ and $n_m$ the refractive index of the scattering particle and the medium respectively):

$$\alpha_{Rayleigh} = 3\varepsilon_0 V_p \frac{(n_s^2 - n_m^2)}{(n_s^2 + 2n_m^2)} \tag{S3}$$

$= 0.128 * 10^{-34}$ and $0.629 * 10^{-34}\ \frac{Cm^2}{V}$.

The scattering cross section is (With $\lambda_{laser,vacuum}$ the laser wavelength:

$$\sigma_{scat} = \varepsilon_0^{-2} \frac{8}{3}\pi^3 \alpha_{Rayleigh}^2 (\lambda_{laser,vacuum}/n_m)^{-4} \tag{S4}$$

$= 0.037 * 10^{-19}$ and $0.814 * 10^{-19}\ m^2$.

And the scattering coefficient ($s$) squared, as the cross section divided by the illuminated area with a diameter of 300 nm:

$$s^2 = \frac{\sigma}{A} = \frac{\sigma_{scat}}{\pi(150 * 10^{-9})^2} \tag{S5}$$

$= 0.047 * 10^{-6}$ and $1.15 * 10^{-6}$.

In the interference term of equation 1 in the main manuscript we can now place the value of $E_{scat} = sE_{inc}$. Here $E_{inc}$ stands for incident field. The second field in the interference term is $E_{ref} = rE_{inc}$ for which we find $r$ (the reflection coefficient) through calculation of the Fresnel reflection coefficient as demonstrated in a paper from the lab of G. Baffou[6]. We just don't use 10 mW but closer to 10 µW of illumination and therefore don't look into the heating effects in our own calculation. We simplify the equation to the version with the field incident normal to the surface which then gives us reflection coefficient $r$. For a single interface (between material 1 and 2) the Fresnel reflection coefficient ($r_{12}$) is[6]:

$$r_{12} = \frac{n_1 - n_2}{n_1 + n_2}. \tag{S6}$$

This can then be substituted into an equation for a 3-layer system, glass (1), Cr (2) and Au (3), which gives[6]:

$$r_{13} = \frac{r_{12} + r_{23} e^{2i\gamma_2 h_2}}{1 + r_{12} r_{23} e^{2i\gamma_2 h_2}}. \tag{S7}$$

With water as the 4[th] and last layer[6] this can go on as:

$$r_{14} = \frac{r_{13} + r_{34} e^{e^{2i\gamma_3 h_3}}}{1 + r_{13} r_{34} e^{2i\gamma_3 h_3}}. \tag{S8}$$

We use the refractive indexes at a wavelength of 633 nm (taken from the website https://refractiveindex.info/?shelf=main&book=Au&page=Johnson, which uses the data from Johnson and Christy 1972) as $n_{glass} = 1.52$, $n_{Cr} = 3.1395 + 3.3152i$, $n_{Au} = 0.18344 + 3.4332i$ and $n_{water} = 1.33$ and for the gold layer $h_3 = 25$ nm thick and $h_2 = 5$ nm for the Cr adhesion layer thickness. Also $\gamma_{layer} = 2\pi(n_{layer}/\lambda_0)$. Now we can do a simplified calculation of the reflection coefficient ($r_{14}$) of our electrode layer interface (glass, Cr, Au, water) and get the following values for modulus and argument of the complex reflection coefficient $r_{14}$. We find the reflection coefficient from modulus $|r_{14}| = 0.58$ and the phase shift $\arg(r_{14}) = -2.13\ rad$.

To double check if these equations that we used make some sense we compared to some trivial cases for reflection. If all materials had only real refractive indices in these equations, the phase shift would be a relatively high positive value which would be as expected from a reflection from a surface of a higher refractive index. Because we now have gold and chromium, which both absorb, the phase shift turns out to be negative because of the complex permittivity (materials that absorb). If in the above equations you put all refractive indices as the same value (index matched values) the phase shift of the reflected light ($E_{ref}$) is zero. Compared to standard mass photometry setups with a simple glass water interface this reflection coefficient is much larger for our Cr and Au layered[6] electrodes than for glass ($r_{glass} = 0.077$ is assumed in mass photometry[4] which is a factor 8 lower). Now we can calculate the interference term with a reflection coefficient of 0.58 as an upper bound since gold is not filling the whole confocal illumination area but rather around half of it. We therefore reduce the reflection coefficient by a factor 2, to *r = 0.29*. Now the interference term for a 20 nm diameter particle becomes:

$$2E_{inc}^2\{\sqrt{s^2} \times r\}\sin(\pi/4) = E_{inc}^2 \times (2 \times sr \times 0.7) = (0.088 \times 10^{-3})I_{inc}, \quad (S9)$$

which is about $(3.0 \times 10^{-4})I_{ref}$. For a 34 nm diameter particle the interference term becomes the following:

$$2E_{inc}^2\{\sqrt{s^2} \times r\}\sin(\pi/4) = E_{inc}^2 \times (2 \times sr \times 0.7) = (4.4 \times 10^{-4})I_{inc}, \quad (S10)$$

which is about $(1.5 \times 10^{-3})I_{ref}$.

**Section S10 Coverslip index mismatch effect on the illumination PSF**

The nanostructure manufacturing obliged us to use thicker coverslips (diced from a larger wafer) with lower index as substrates than commonly used in optical microscopy. This deviation from normal optics had direct consequences on the shape and the intensity of the illumination PSF produced with our high-NA objective. High-quality infinity-corrected objectives are corrected for a specific working distance and in the case of oil-immersion objectives, each interface (i.e. objective lens/oil/coverslip) must match the designed refraction indices (± 0.002).

Figure S8 shows a simulation of the effect on the illumination PSF of a refractive index mismatch between the immersion oil and the glass coverslip in combination with a thicker coverslip. The interface between the water and the glass is at $Z=0$ and all the negative values of Zare in the glass when the positives ones are in water. Figure S8A shows the illumination PSF for the perfect case where the immersion oil matches the objective lens and the coverslip index. In that situation, the PSF shows a transverse- FWHM of 280 nm and an axial FWHM of 512 nm. Figure S8B presents a simulation for an index mismatch ($n\_oil=$ 1.518 and $n\_glass=$ 1.480) and a thicker coverslip (200 µm). In that case, the PSF has a transverse FWHM of 319 nm and an axial FWHM of 1.104 nm. In the case of the index mismatch and a thicker coverslip, we can observe a larger PSF in X direction but also several other local extremums, positioned further in the water (i.e. away from the surface).

To understand the situation from a geometrical point of view, Fig. S8(C) shows the ray tracing coming into the objective and focusing on the sample for optimal objective working conditions as it is simulated for (A). Figure S8(D) represents the ray tracing for refractive index mismatch and a thicker coverslip (dashed line: case (C)) where we can see the position of the illumination PSF being shifted into the sample.

In practice, the focus is found by adjusting the Z position of the objective to find the maximum signal received from a point source (luminescent bead for example). The collection PSF is then not overlapping with the illumination PSF and leads to a loss of resolution as the effective NA gets lower (i.e. high angle ray are not collected from the new focus position). The consequence for an experiment is to deposit less energy on the area of interest, with a loss of resolution. It is shown in section S6] that the fluorescent signal collected from spin coated particles in the case of a good index match is higher than for the same particles trapped with the sample using thick and different index coverslips. Another aspect to consider is the shadowing effect of the electrodes over the excitation beam that could reduce the amount of light received by the particles.

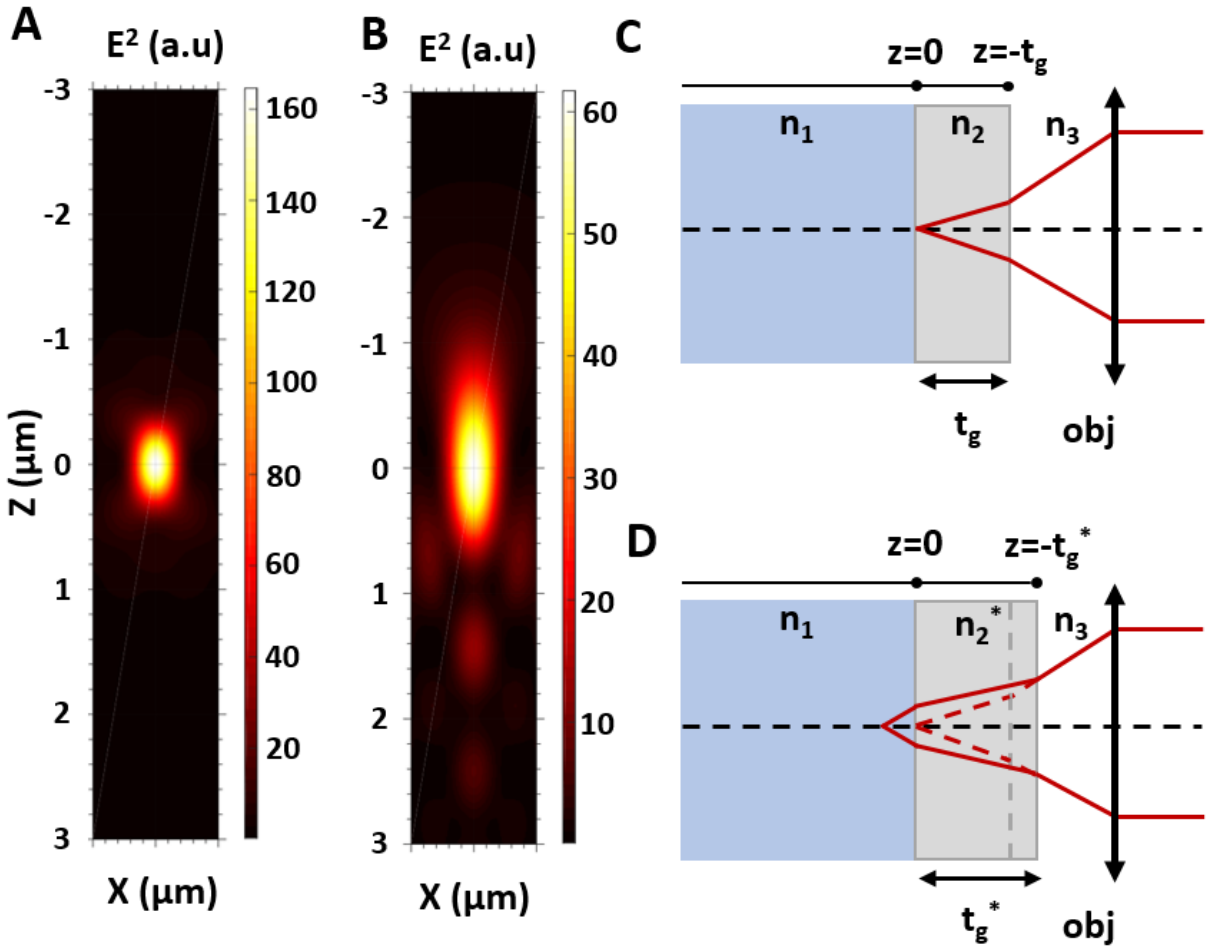

Figure S8: Simulated illumination PSF for an objective x60 NA 1.4 and a wavelength of 632 nm, (A) for the perfect case of index match between the coverslip and the immersion oil ($n_2=n_3$). (B) Simulation for an index mismatch ($n_2=1.480$ for the coverslip and $n_3=1.518$ for the immersion oil) and for an extra thickness of the coverslip (200 µm instead of 170 µm). The illumination path of a confocal microscope for (C) the perfect case simulated on (A), (D) situation where an index mismatch and a thicker coverslip produce a shift of the illumination and collection PSF. Simulation and design with the open-source software PSFLab[7] developed by Michael J. Nasse, Jörg C. Woehl.

**Section S11 Some examples of the events with modulation of the trapping potential**

In addition to the one single event that is showed and analyzed in Figure 5 of the main text we measured additional events at 500 Hz modulation during the same experiment. Moreover, we have trapped particles at different modulation frequencies, $f_m$, from 400 to 1000Hz. Figure S9 contains 21 events registered at different modulation frequencies. For every event we plot 4 panels in Figure S9: fluorescence time trace (top left, dark red), scattering time trace (bottom left, light gray for 1MS/s sampling, no filtering , and red for line filtering at the modulation frequency), power spectral density of the unfiltered scattering signal when the particle is trapped (top right, red), and power spectral density of the signal applied to the electrodes (bottom right, blue). For every modulation frequency from 400 to 1000 Hz we have identified two events. The power spectral density of the scattered signal contains the modulation frequency and using a digital lock-in to this frequency allows to extract the signal from the noise. The spectra of the signal applied to the electrodes contains the modulation component around 100kHz, which contains 3 peaks: 100kHz, 100kHz + $f_m$, 100kHz – $f_m$ as expected for the AM modulation. The low-frequency part of this spectra contains frequencies that come through electrical interferences in our setup. Those peaks are at 188 Hz, 540 Hz, 635 Hz, 1081 Hz, 1621 Hz, and a tiny one at 2103 Hz (Figure S10). They are present in the spectra of the electrode voltage of every event but these peaks don't make it into the scattering spectra. Scattering spectra only features the modulation frequency, sometimes accompanied by higher harmonics.

Interestingly, the Event 4, which is modulated at 500Hz we see that we trap some object that is not fluorescent first, and only then it is followed by a fluorescent nanoparticle.

Event 1: Modulation Frequency 500 Hz

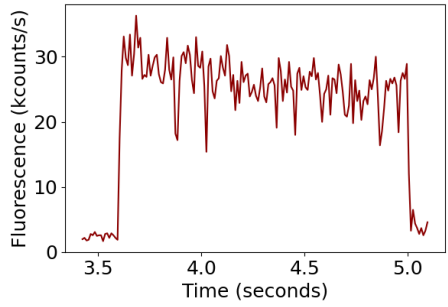
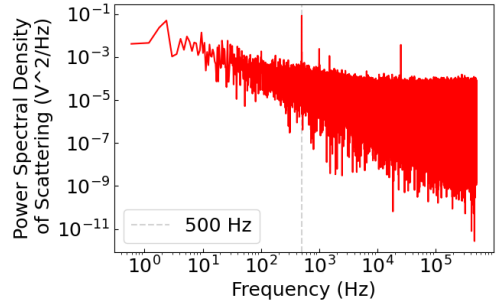
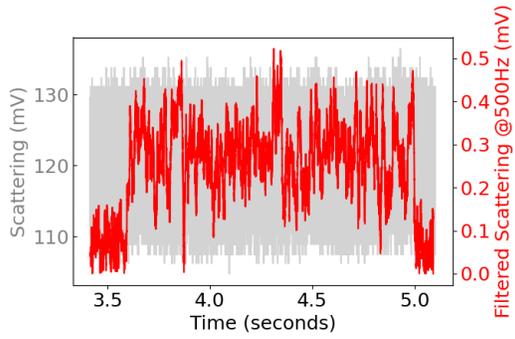
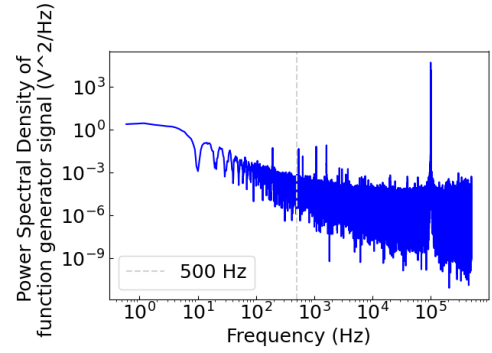

Event 2: Modulation Frequency 500 Hz

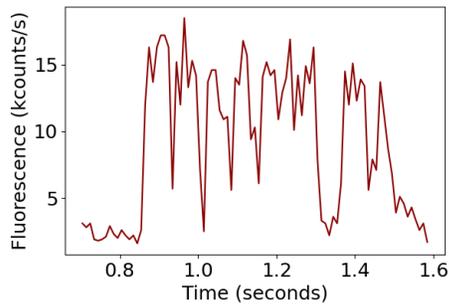
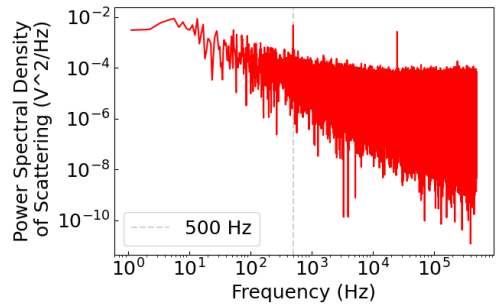
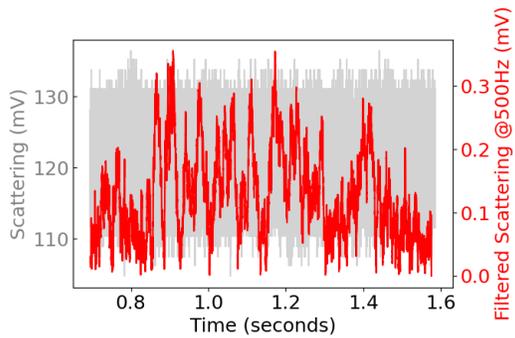
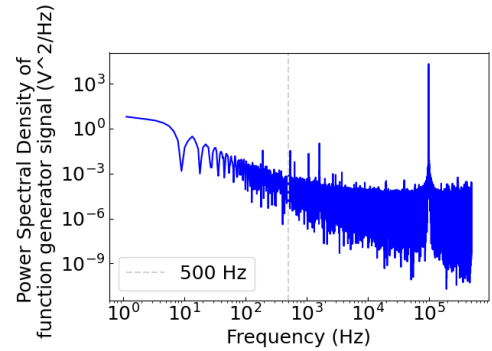

Event 3: Modulation Frequency 500 Hz

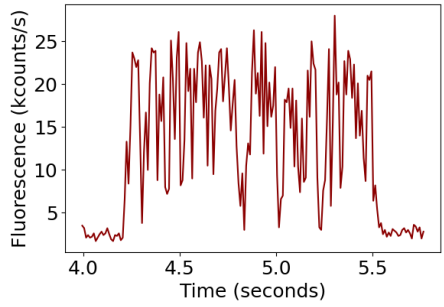
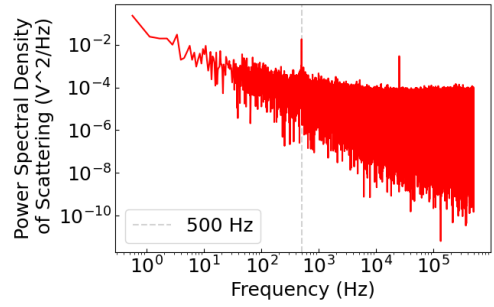
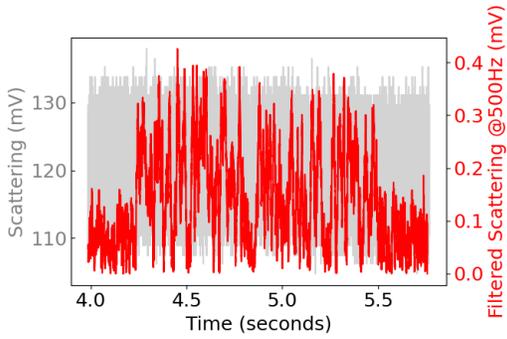
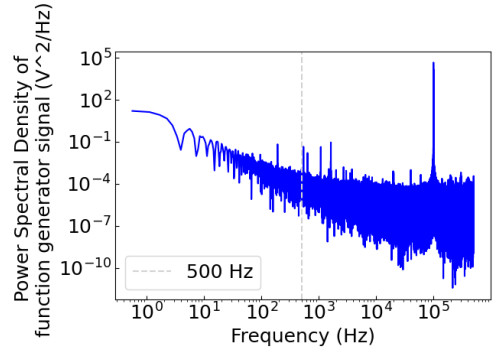

Event 4: Modulation Frequency 500 Hz

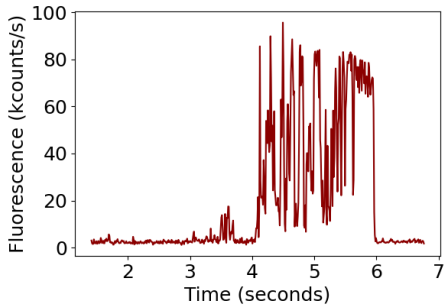
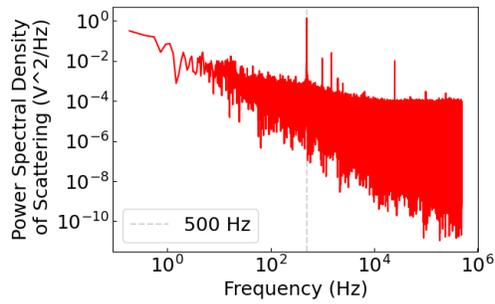
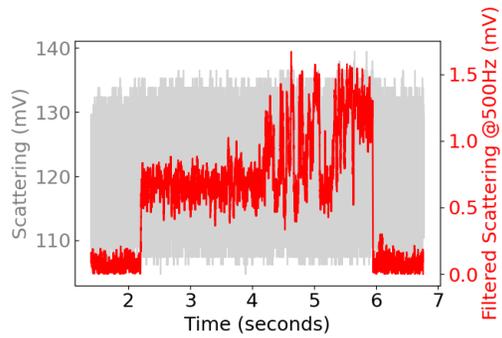
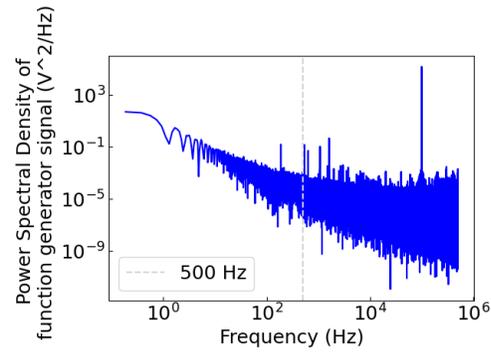

### Event 5: Modulation Frequency 500 Hz

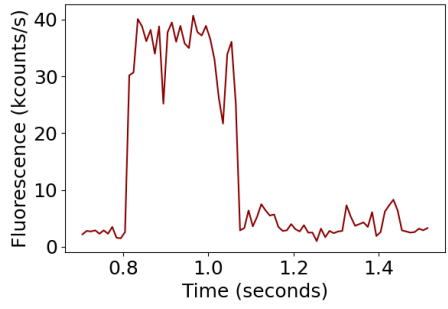
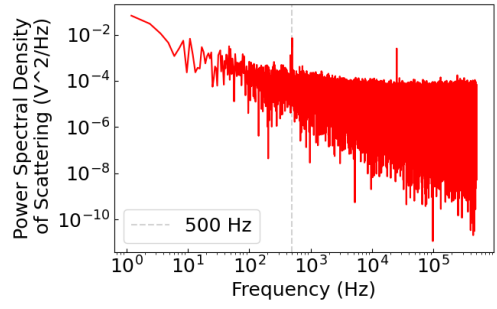
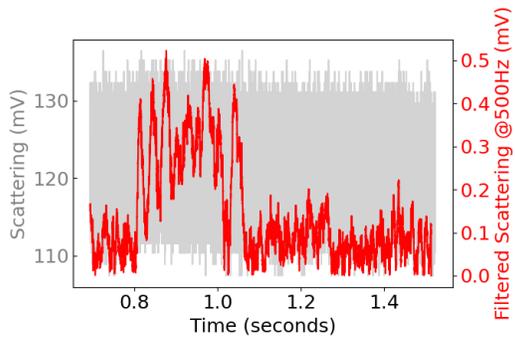
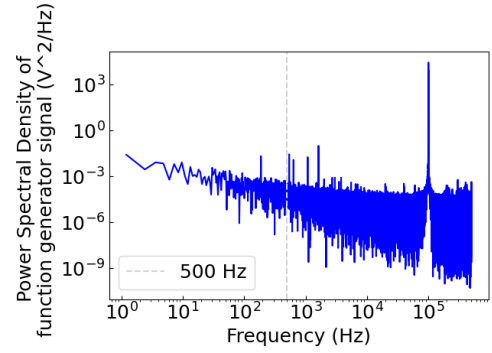

### Event 6: Modulation Frequency 500 Hz

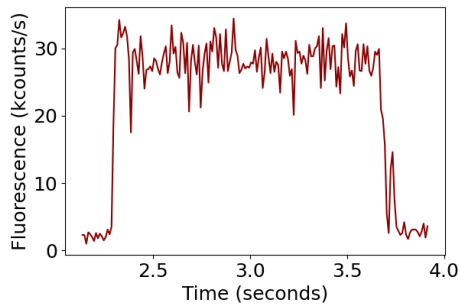
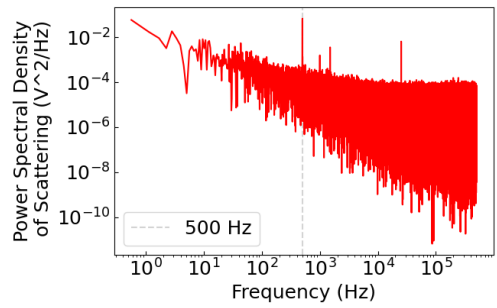
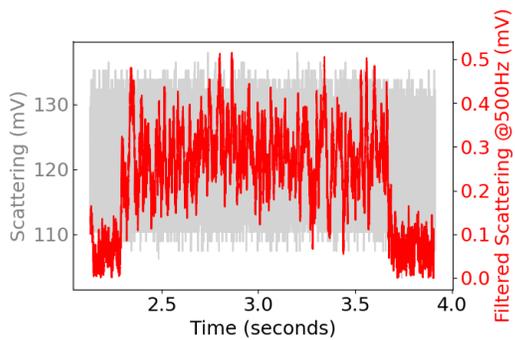
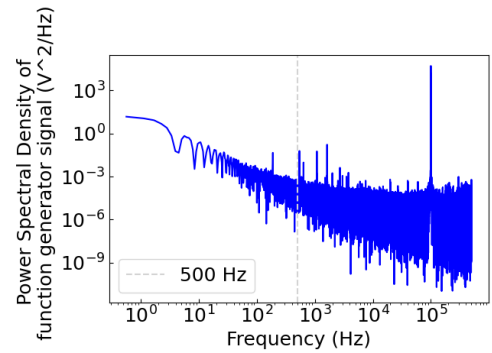

Event 7: Modulation Frequency 1000 Hz

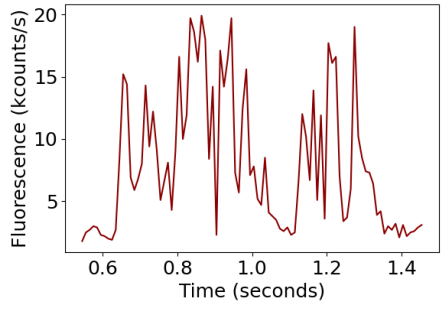
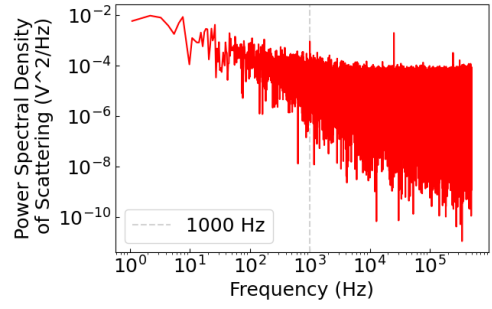
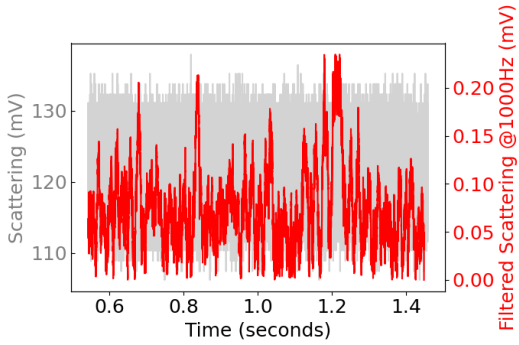
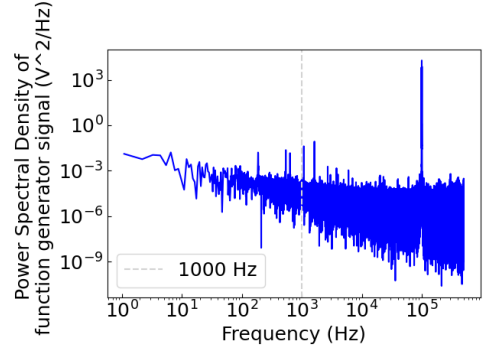

Event 8: Modulation Frequency 1000 Hz

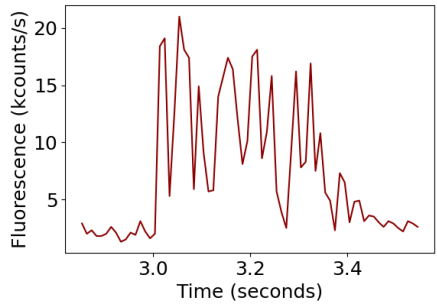
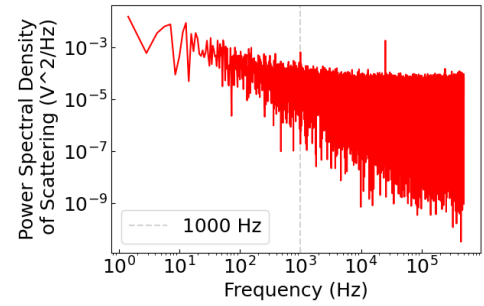
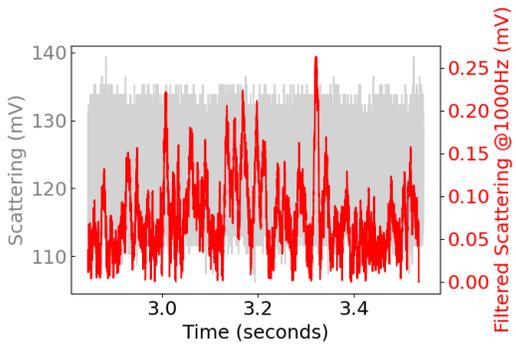
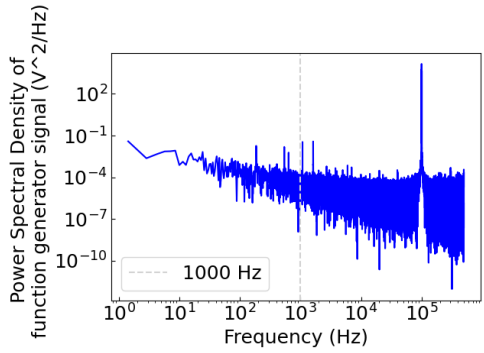

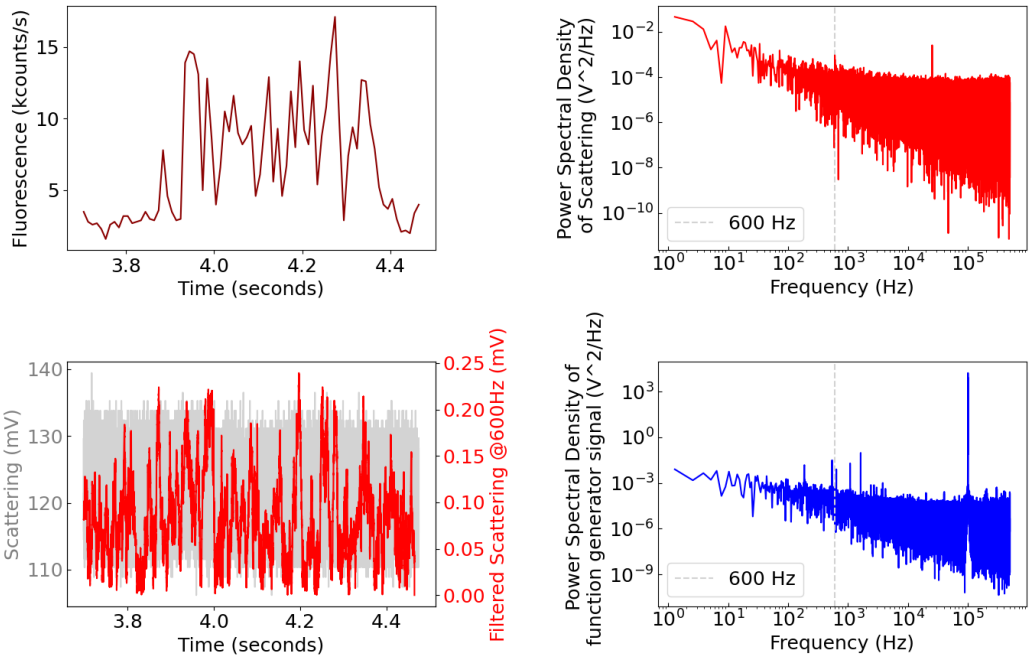
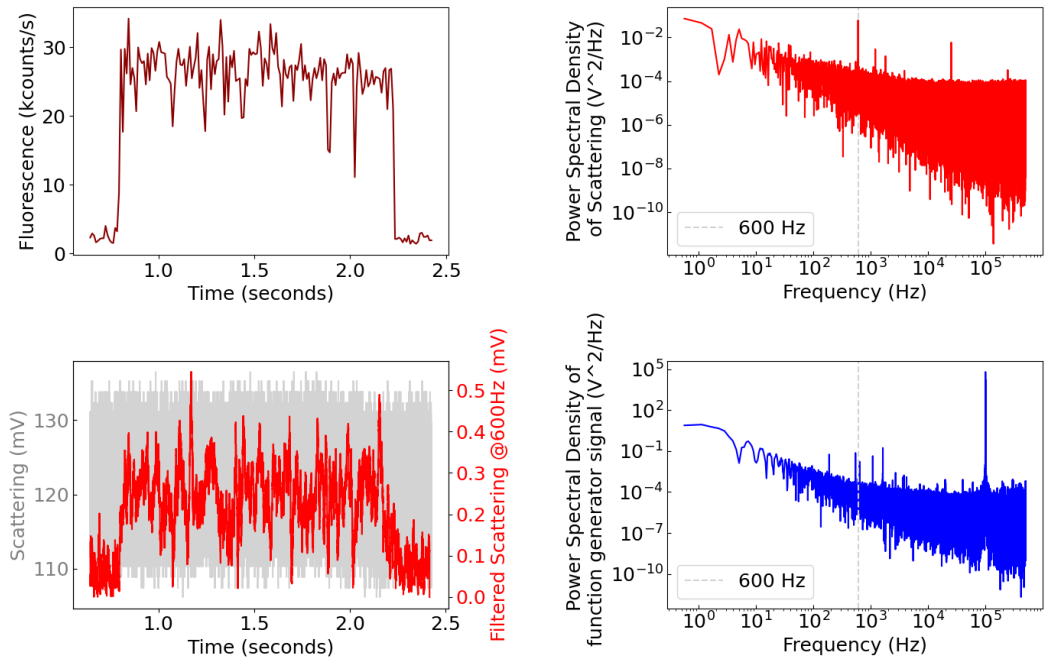

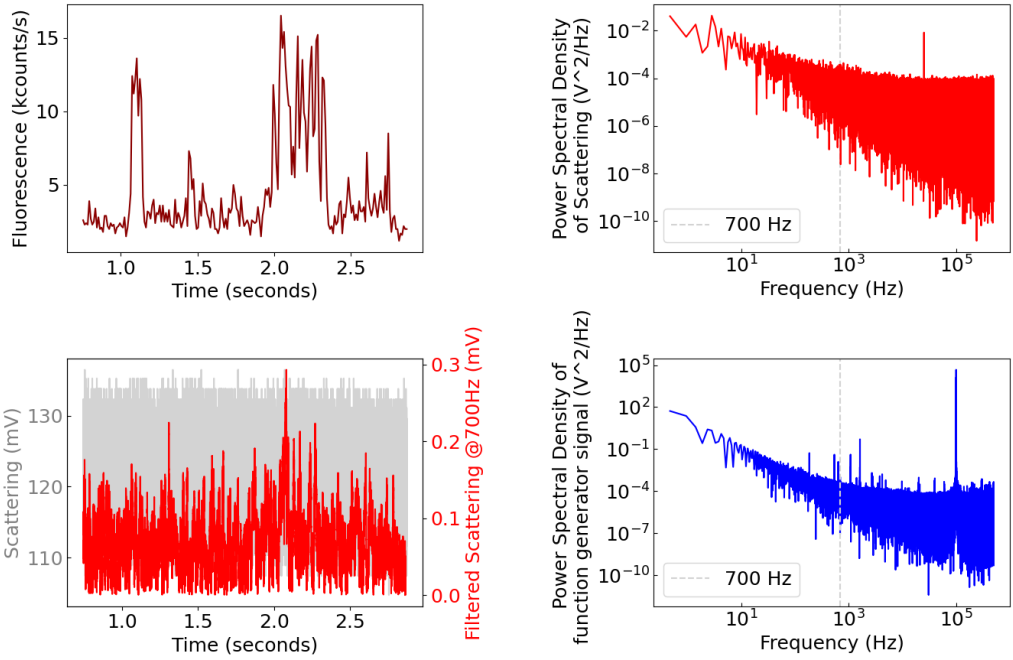

Event 11: Modulation Frequency 700 Hz

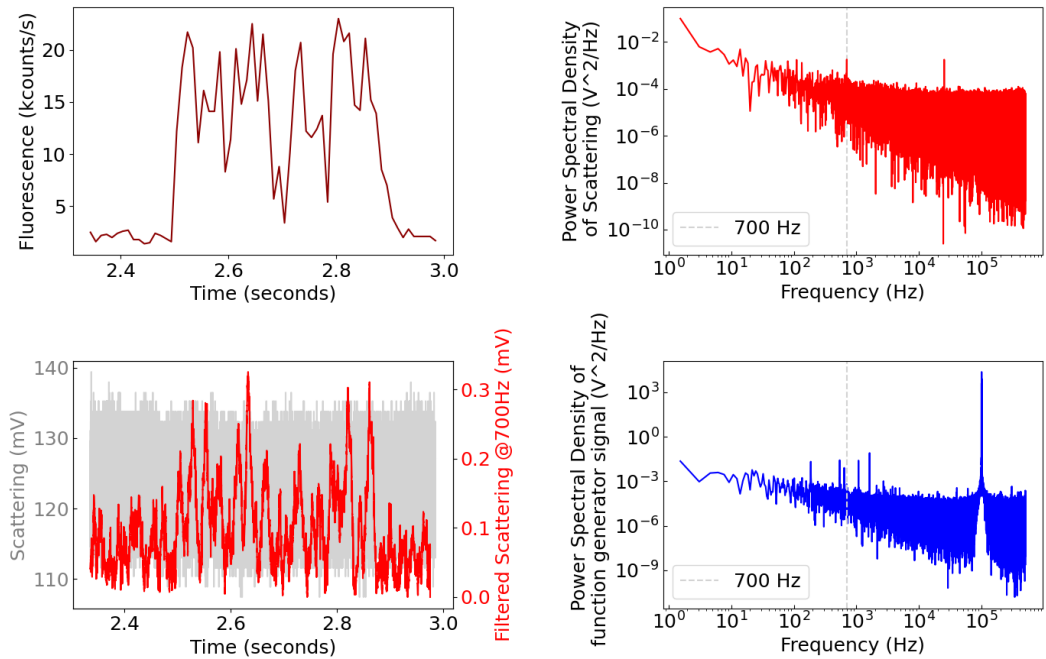

Event 12: Modulation Frequency 700 Hz

Event 13: Modulation Frequency 800 Hz

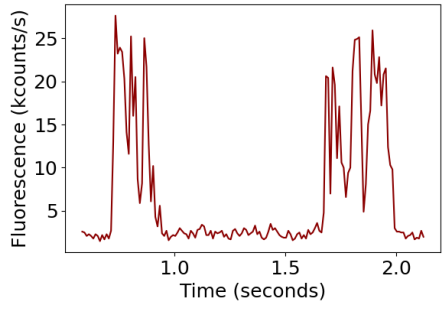
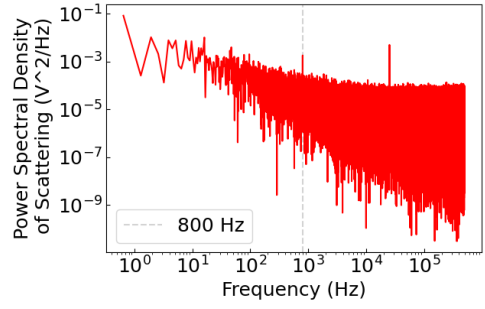
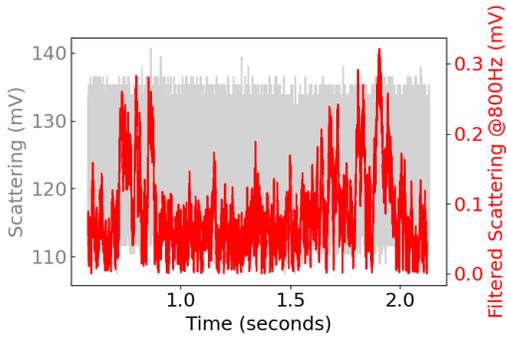
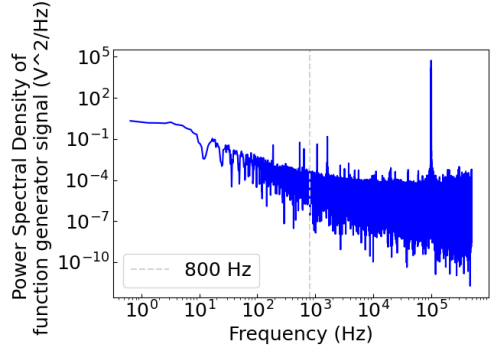

Event 14: Modulation Frequency 800 Hz

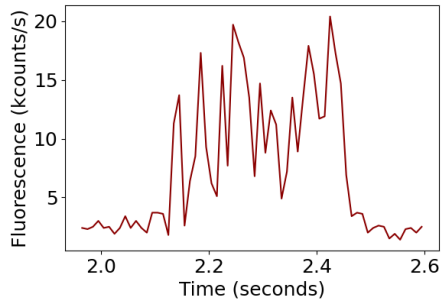
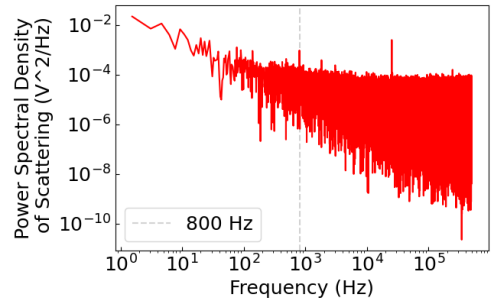
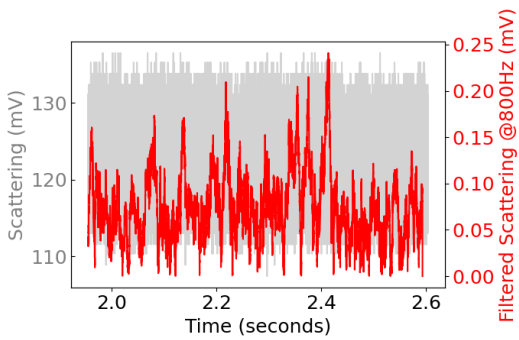
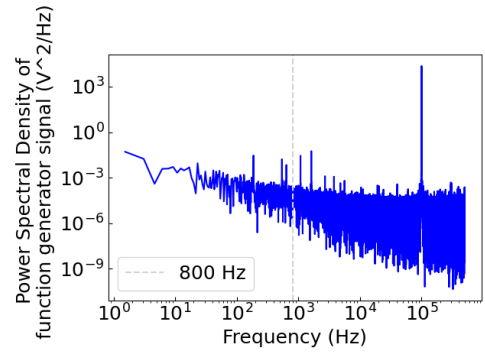

Event 15: Modulation Frequency 800 Hz

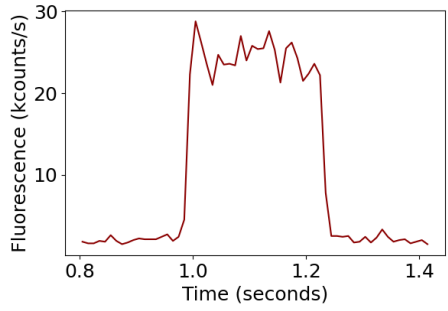
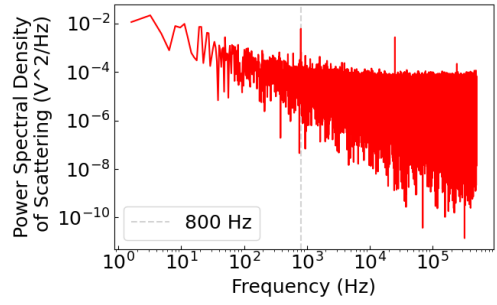
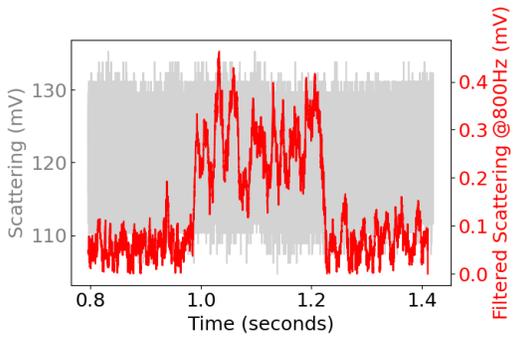
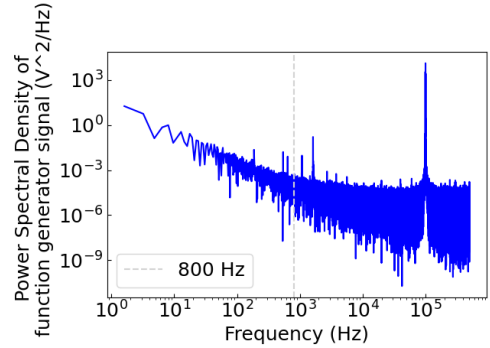

Event 16: Modulation Frequency 800 Hz

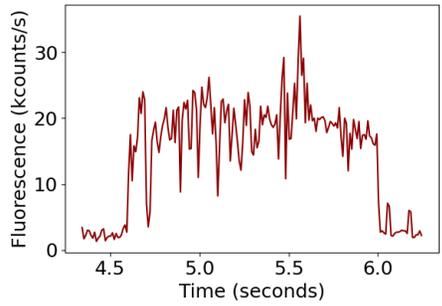
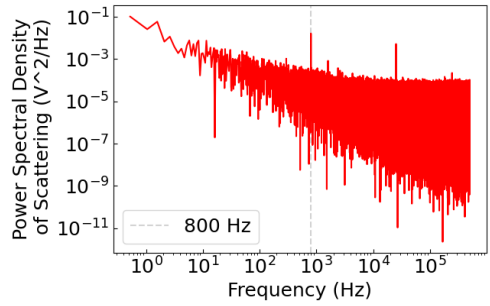
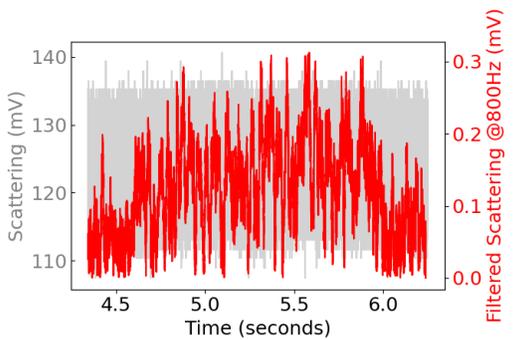
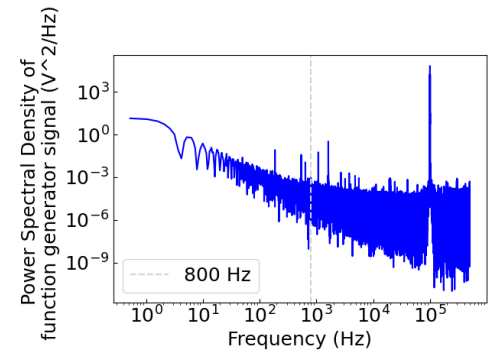

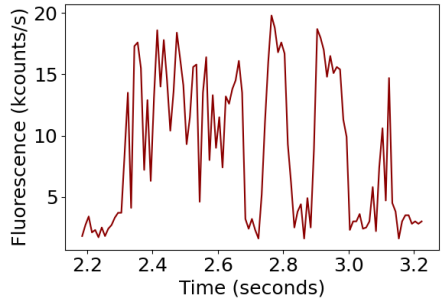
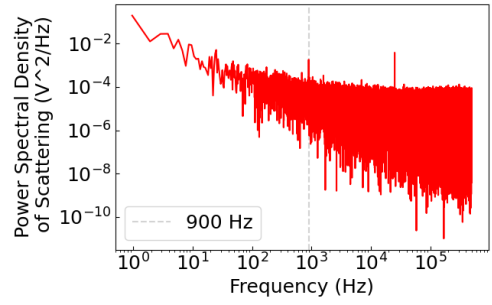
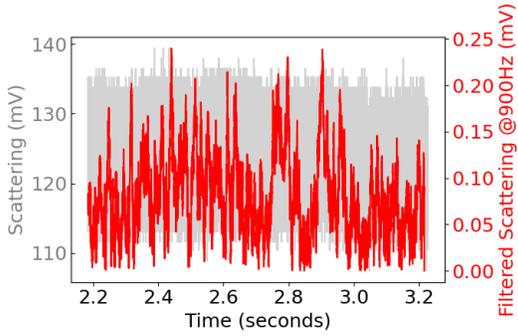
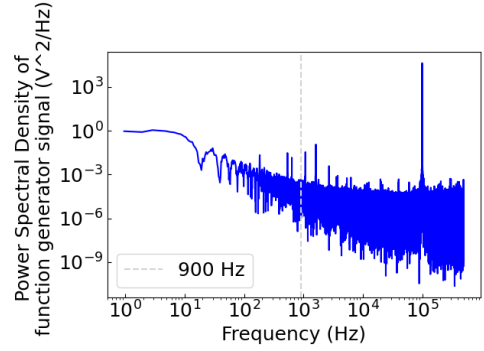
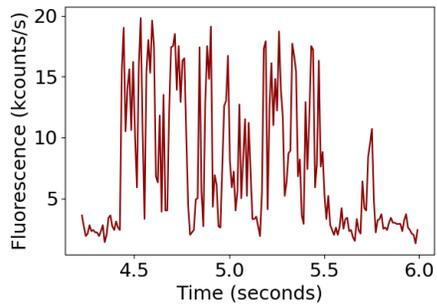
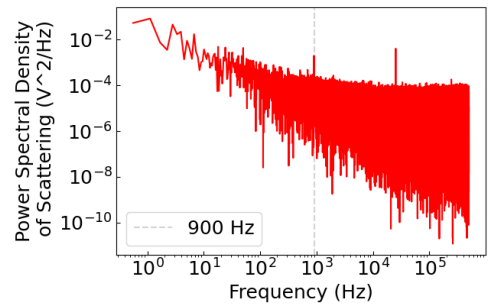
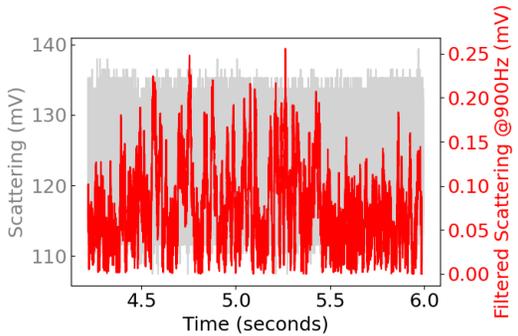
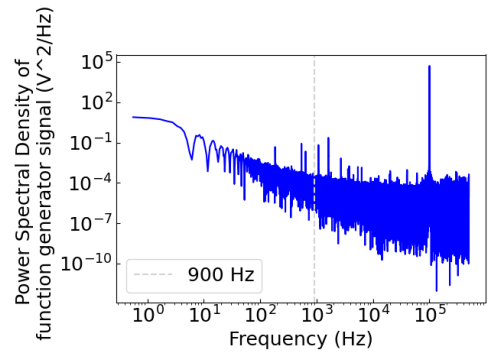

Event 19: Modulation Frequency 900 Hz

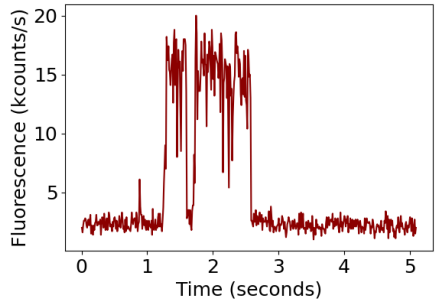
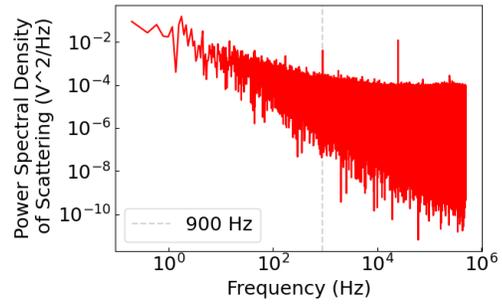
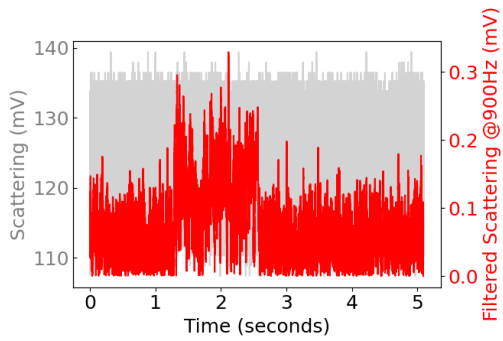
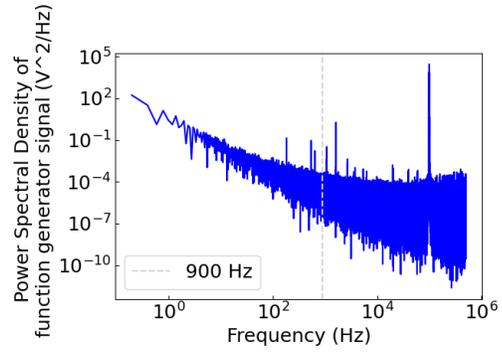

Event 20: Modulation Frequency 400 Hz

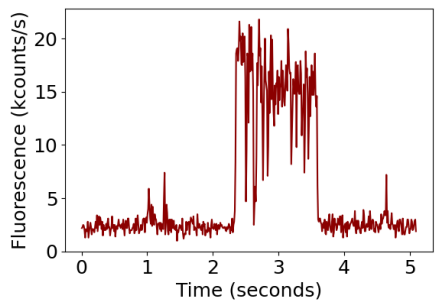
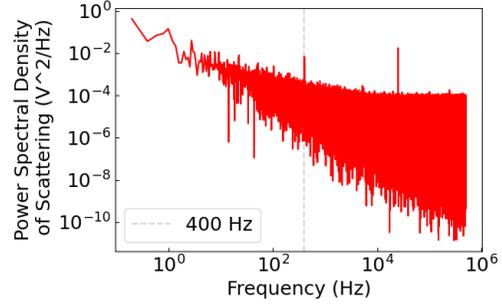
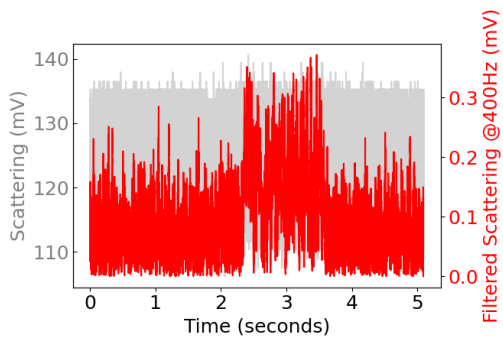
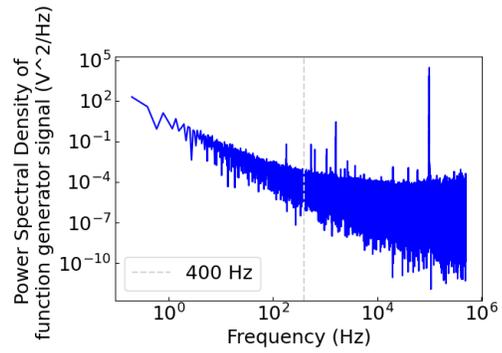

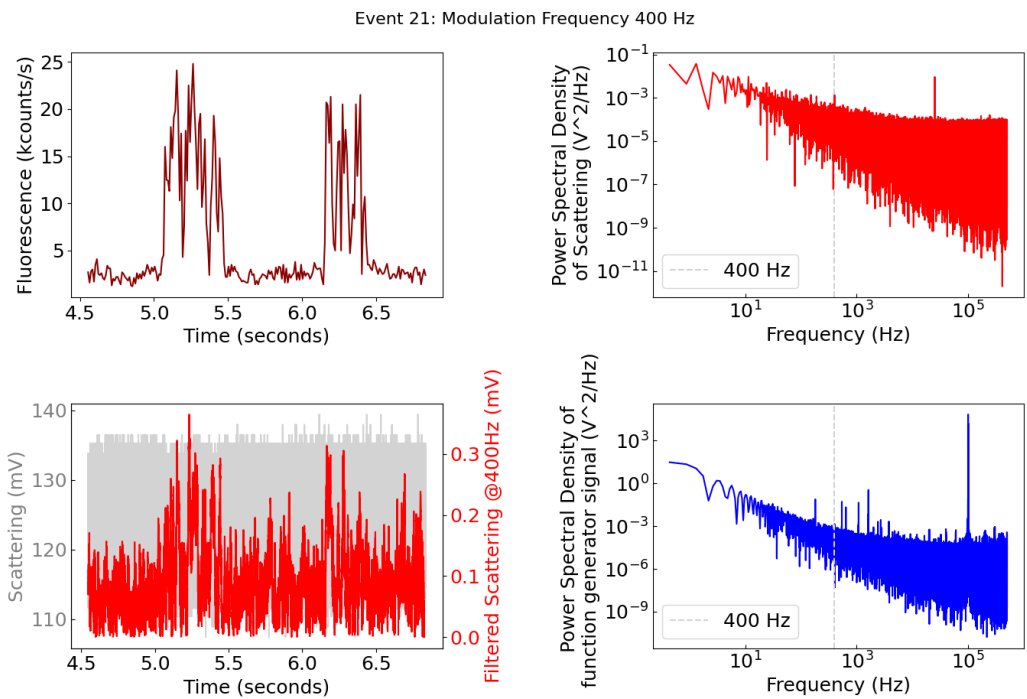

Figure S9 Events featuring actuation of the trapped nanoparticle by modulation of the trapping potential. The figure contains 21 4-panel subfigures each dedicated to a single event. The integration time for the fluorescent signal and for the line filtering in are 10 ms. Every subfigure features a fluorescence time trace (top left, dark red), a scattering time trace (bottom left, light gray for 1MS/s sampling, no filtering , and red for line filtering at the modulation frequency), a power spectral density of the unfiltered scattering signal when the particle is trapped (top right, red), and a power spectral density of the signal applied to the electrodes (bottom right, blue). All the spectra have the modulation frequency indicated with a light grey vertical line.

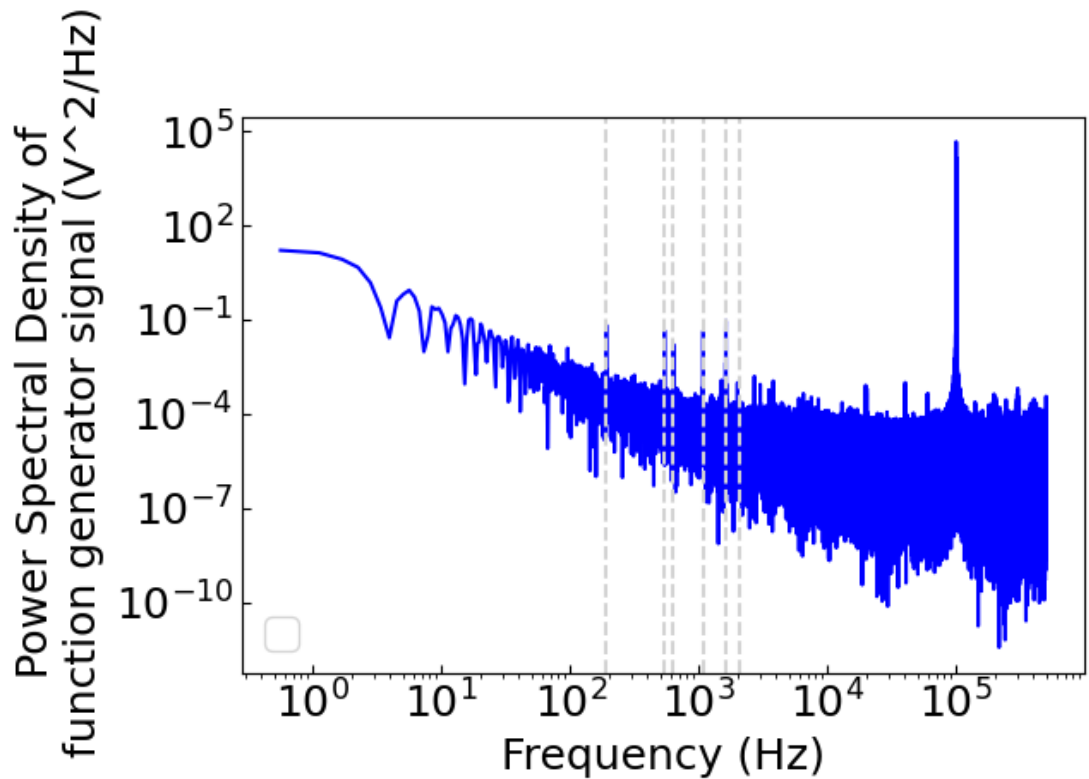

Figure S10 Power spectral density of the modulated signal applied to the electrodes. The vertical dotted lines mark the electrical interferences at 188 Hz, 540 Hz, 635 Hz, 1081 Hz, 1621 Hz, and a tiny one at 2103 Hz

**Section S12: Tween20**

We used a surfactant to make sure the particles wouldn't permanently stick to the gold surface after being trapped. This sticking was a major hurdle during our initial experiments.

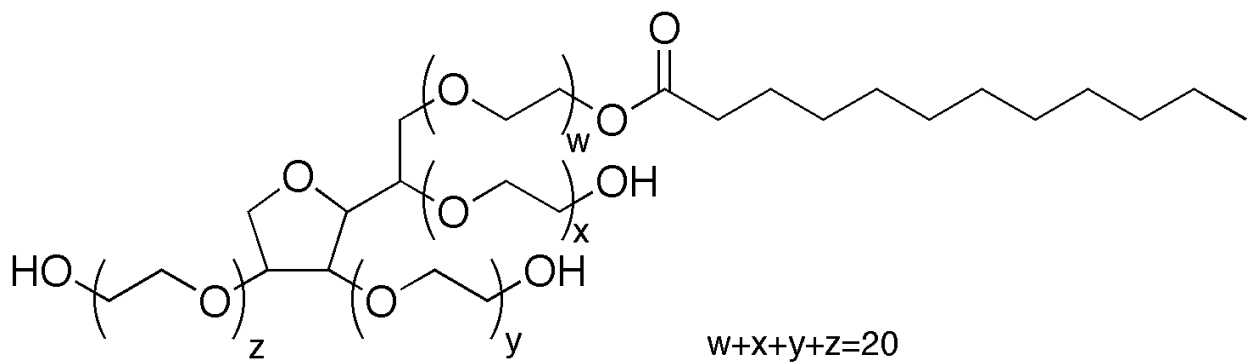

Figure S13: Surfactant TWEEN 20 (Polysorbate 20, nonionic ) as found on the websites of Sigma Aldrich (https://www.sigmaaldrich.com/NL/en/product/sial/p1379, accessed 3 May 2024)

**Section 13: Optical setup details**

The layout of the optical setup is shown in figure S14. It is a confocal microscope in reflection geometry and the beam path is designed such that it maintains the polarization[8] in both the illumination and the detection path, by keeping the incidence angles on crucial mirrors and beam splitters small (~5 degree). The linearly polarized output beam of a helium-neon laser ($\lambda$ = 633 nm) is spectrally cleaned by a 633 nm clean-up filter, then linearly polarized by a fixed-orientation Glan-Thompson prism (which in combination with a rotatable ½ wave plate at the laser output is used to adjust the illumination beam intensity) after which the polarization is manipulated by a combination of a ½ wave plate and a ¼ wave plate (Halle), when working in circularly polarized mode. The ¼ wave plate is removable to make any linear polarization angle possible. After this the beam is expanded and spatially filtered by sending the light through a combination of a $f$=40 mm lens, a 75 µm pinhole and a $f$ =200 mm lens (both doublets). The beam goes through an iris opened as wide as the width of the back entrance of the objective. The beam is then reflected by a 70(T):30(R) beam-splitter (Thorlabs) and two silver mirrors (Thorlabs), that are orthogonal to each other, pointing the beam upwards to the back entrance of the objective (Olympus 60x, NA 1.4). The objective is slightly overfilled and the power used during the experiments was close to 10 µW before entering the objective. The laser power is adjusted so that the Si-PIN photodiode detector (Femto) gives an output of close to 1 V to stay in linear response regime of its gain setting $10^7$. The laser power is adjusted for each nano-gap to compensate for fabrication variability, which influences the reflectivity of the substrate, and the photodiode detector output is kept constant for every experiment. Light from the sample is collected by the same objective through which it was illuminated. Then the light is transmitted through the beam-splitter (70T:30R) into the detection path. The light coming from the objective is separated into its scattering/reflection component and the fluorescence component by a dichroic mirror (LP > 650 nm, Semrock). Then the scattered/reflected light is focused onto a confocal pinhole with the Si-PIN photodiode detector behind it. Before the detector there is a 633 nm clean-up line filter (Semrock) to block light from luminescence. Some experimental calibration steps required a polarization analyzer in the scattering detection path in the form of a linear polarization filter. Fluorescence and also some part of the light emitted by the gold film goes through the dichroic mirror and is directed through a set of emission filters (Band pass 653-712 nm) to a single-photon-counting module (SPCM, PerkinElmer) through a 75 µm pinhole.

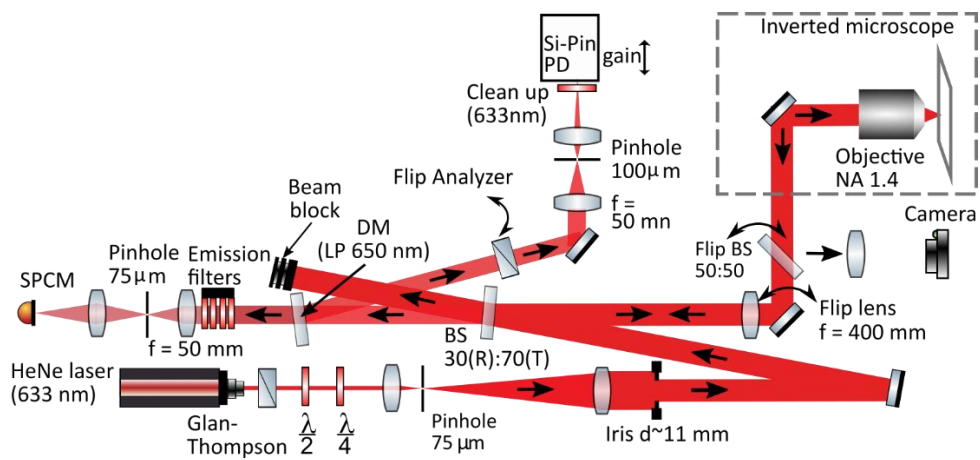

Figure S14: Schematic of the optical setup. BS = beam-splitter, Clean up = Clean up narrow line filter, DM = Dichroic mirror, SPCM = single-photon counting module; the detector has an adjustable gain setting. Only lenses being part of a spatial filter have been indicated.

To be able to find the region of interest where the electrode tip is located, a flip beam splitter (Thorlabs, pellicle 50:50) and a flip 400 mm lens is added to the setup. In this configuration the wide field light is collected by a second 400 mm lens and focused onto a CMOS camera (Photometrics, CoolSNAP EZ).

**Supplementary Sections References:**